\documentclass[letterpaper,11pt]{article}
\pdfoutput=1 
\usepackage{jheppub2} 
\usepackage{bbm,bm,graphicx,mathtools,color,hyperref,slashed}
\usepackage{amssymb,amsmath}
\usepackage[T1]{fontenc}
\usepackage{enumerate}

\usepackage[all]{xy}

\graphicspath{{./Figures/}}

\newcommand{\diff}{\mathrm{d}}
\newcommand{\p}{\partial}
\newcommand{\ve}{\varepsilon}

\newcommand{\be}{\begin{equation}}      
\newcommand{\ee}{\end{equation}}      
\newcommand{\bea}{\begin{eqnarray}}      
\newcommand{\eea}{\end{eqnarray}}

\newcommand{\im}{\mathrm{i}}

\newcommand{\calZ}{\mathcal{Z}}

\newcommand{\rme}{\mathrm{e}}

\newcommand{\A}{X}

\begin{document}

\title{Lattice gauge theory for Haldane conjecture and central-branch Wilson fermion}

\author[1,2,3]{Tatsuhiro Misumi,}
\emailAdd{misumi@phys.akita-u.ac.jp}

\author[4]{Yuya Tanizaki}
\emailAdd{ytaniza@ncsu.edu}

\affiliation[1]{Department of Mathematical Science, Akita University, Akita 010-8502, Japan}
\affiliation[2]{iTHEMS Program, RIKEN, Wako 351-0198, Japan}
\affiliation[3]{Research and Education Center for Natural Sciences, Keio University, Kanagawa 223-8521, Japan}
\affiliation[4]{Department of Physics, North Carolina State University, Raleigh, NC 27607, USA}

\abstract{
We develop the $(1+1)$d lattice $U(1)$ gauge theory in order to define $2$-flavor massless Schwinger model, and discuss its connection with Haldane conjecture. We propose to use the central-branch Wilson fermion, which is defined by relating the mass, $m$, and the Wilson parameter, $r$, as $m+2r=0$.  This setup gives two massless Dirac fermions in the continuum limit, and it turns out that no fine-tuning of $m$ is required because the extra $U(1)$ symmetry at the central branch, $U(1)_{\overline{V}}$, prohibits the additive mass renormalization. Moreover, we show that Dirac determinant is positive semi-definite and this formulation is free from the sign problem, so the Monte Carlo simulation of the path integral is possible. By identifying the symmetry at low energy, we show that this lattice model has the mixed 't~Hooft anomaly between $U(1)_{\overline{V}}$, lattice translation, and lattice rotation. We discuss its relation to the anomaly of half-integer anti-ferromagnetic spin chains, so our lattice gauge theory is suitable for numerical simulation of Haldane conjecture. Furthermore, it gives new and strict understanding on parity-broken phase (Aoki phase) of $2$d Wilson fermion.
}

\maketitle
\section{Introduction}\label{sec:introduction}

Quantum field theory (QFT) provides us a useful description about low-energy behaviors of quantum spin chains.  
A seminal work by Haldane has shown that anti-ferromagnetic spin chain can be described by the $(1+1)$d relativistic $S^2$ sigma model with a topological term, and this leads to a striking conclusion that the system is gapless for half-integer spins while it is gapped for integer spins~\cite{Haldane:1982rj, Haldane:1983ru}. 
It is very interesting that we can explain why the half-integer spin chains cannot have the unique gapped vacuum by the Lieb-Schultz-Mattis (LSM) theorem~\cite{Lieb:1961fr, Affleck:1986pq, PhysRevLett.84.1535}, and this theorem allows only two possible low-energy behaviors: gapless phase, and dimerized phase. 
Recently, it is understood that the LSM theorem is essentially identical to the 't~Hooft anomaly matching~\cite{tHooft:1979rat, Frishman:1980dq} thanks to their connection with symmetry-protected topological (SPT) order via boundary-bulk correspondence~\cite{Wen:2013oza, Kapustin:2014lwa, Cho:2014jfa, Wang:2014pma}. 
This brings us a renewed attention to anomaly matching condition, and lots of new aspects of nonperturbative physics are discovered~\cite{ Witten:2016cio, Tachikawa:2016cha, Gaiotto:2017yup, Tanizaki:2017bam, Kikuchi:2017pcp, Komargodski:2017dmc, Komargodski:2017smk, Shimizu:2017asf, Wang:2017loc,Gaiotto:2017tne, Tanizaki:2017qhf, Tanizaki:2017mtm,  Yamazaki:2017dra, Guo:2017xex,  Sulejmanpasic:2018upi, Tanizaki:2018xto, Yao:2018kel, Kobayashi:2018yuk,  Tanizaki:2018wtg,Anber:2018jdf,   Anber:2018xek, Armoni:2018bga, Hongo:2018rpy,  Yonekura:2019vyz, Nishimura:2019umw, Misumi:2019dwq, Cherman:2019hbq}. 

Quite often, the QFT descriptions of spin systems are strongly coupled. In order to go beyond the kinematical constraints about its low-energy properties, we have to develop first-principle numerical computations of those systems. 
Lattice gauge theory~\cite{Wilson:1974sk, Creutz:1980zw}  is one of the most reliable techniques to study non-perturbative physics of asymptotically-free QFTs, including Yang-Mills theory and Quantum Chromodynamics (QCD).
However, the formulation of lattice gauge theory with fermions is not straightforward due to the difficulty of realization of a single chiral-symmetric fermion, which is naively forbidden by the Nielsen-Ninomiya theorem~\cite{Karsten:1980wd, Nielsen:1980rz, Nielsen:1981xu, Nielsen:1981hk}.  
Moreover, the numerical cost of simulations mainly originates in the calculation of quark determinant. 
These facts indicate that the fermion discretization is a key to efficient lattice simulations of quantum field theories.

Among the several lattice fermion formulations, Wilson fermion~\cite{Wilson:1975id} has been used broadly. Although explicit chiral symmetry breaking in the formulation leads to additive mass renormalization, it successfully describes QCD by fine-tuning of the mass parameter. On the other hand, the recent progress of understanding of topological insulators and SPT orders sheds a light on another aspect of Wilson fermion:
the Wilson fermion with negative mass parameter, which contains only massive degrees of freedom, corresponds to nontrivial SPT phase, where the transition to another SPT phase requires the gap to be closed once.
This viewpoint clearly exhibits that a massless fermion, or a gapless mode, appears at the boundary between theories in different SPT phases, which is nothing but the chiral-symmetric lattice fermion formulation called Domain-wall fermion~\cite{Kaplan:1992bt, Shamir:1993zy, Ginsparg:1981bj, Neuberger:1998wv}.

It is known that the Wilson term breaks the whole $U(4)\times U(4)$ flavor and chiral symmetries of naive lattice fermion to a single $U(1)$ vector symmetry.
However, it was shown that the Wilson fermion with the special mass parameter $m+d r=0$ has larger symmetry as $U(1)\times U(1)$ \cite{Creutz:2011cd, Kimura:2011ik}, where $d$ stands for the spacetime dimensions and $r$ is the Wilson parameter.
This case is called a ``central-branch Wilson fermion'', whose possibility of being applied to QCD simulations has been discussed in terms of strong-coupling expansion~\cite{Kimura:2011ik}, the Gross-Neveu model~\cite{Creutz:2011cd} and the lattice perturbation~\cite{Misumi:2012eh,Chowdhury:2013ux}.
It is notable that the proper choice of flavored-mass terms (generalized Wilson terms)~\cite{Creutz:2010bm, M1} leads to two-flavor central-branch Wilson fermion while the usual Wilson fermion has six species on the central branch in 4d.

In this work, we focus on the 2d central-branch Wilson fermion, which produces two massless degrees of freedom, with emphasis on its physics near the continuum limit.
Based on the 2d central-branch Wilson fermion, we define $2$-flavor massless Schwinger model and discuss its connection with Haldane conjecture. 
We identify its symmetry at low energy scale and find that this lattice model has the same 't~Hooft anomaly with that of half-integer spin chain. 
We emphasize that all the symmetries relevant for 't~Hooft anomaly are exact symmetries at the lattice level, and this means that we find the $\mathbb{Z}_2$ 't~Hooft anomaly between $U(1)\times U(1)$ symmetry at the central branch, lattice translation, and lattice rotation symmetries. 
Moreover, we show that this setup is free from the sign problem since the Dirac determinant is positive semi-definite, thus we can perform the Monte Carlo simulation of the system in principle. 
We also discuss implications of our study on the parity-broken phase (Aoki phase) of Wilson fermion.

\section{Central-branch Wilson fermion}
\label{sec:Wil}

In this section, we first give a brief review on Wilson fermion and central-branch Wilson fermion.
We begin with looking into flavor-chiral symmetry of naive fermions by following Ref.~\cite{Kimura:2011ik}. 
After that, using this knowledge, we discuss the symmetry of the Dirac spectrum for the central-branch Wilson fermion. 
Using this symmetry, we prove that the Dirac determinant of the central-branch Wilson fermion is positive definite on even-sites lattice. 
This shows that the numerical Monte Carlo simulation is possible. 

\subsection{Wilson fermion and central branch}

The 2d Wilson fermion action is
\begin{equation}
S_{\rm W} =\underbrace{\sum_{n}\sum_{\mu=1,2}\overline{\psi}_{n}\gamma_{\mu}D_{\mu}\psi_{n}}_{\text{naive kinetic term}}
\,+\, \underbrace{\sum_{n}m\overline{\psi}_{n}\psi_{n}}_{\text{mass term}}
+\,\underbrace{r \sum_{n}\sum_{\mu=1,2}\overline{\psi}_{n}(1-C_{\mu}){\psi}_{n}}_{\text{Wilson term}},
\label{WilS}
\end{equation}
where $D_{\mu}\equiv(T_{+\mu}-T_{-\mu})/2$, 
$C_{\mu}\equiv (T_{+\mu}+T_{-\mu})/2$ with $T_{\pm\mu}\psi_{n}=U_{n,\pm\mu}\psi_{n\pm\mu}$, respectively. 
The sum, $\sum_n$, stands for the summation over spacetime lattice sites, $n=(x,y)\in \mathbb{Z}\times \mathbb{Z}$. 
Because of the Wilson term, the degeneracy of four species in naive fermion is lifted into three branches, 
where one, two and one flavors lives.

The 2d massless naive action possesses $U(2)\times U(2)$ flavor-chiral symmetries, which is a remnant of the whole flavor-chiral symmetry of 4 species. (See \cite{Kimura:2011ik} for the symmetries in 4d.)
The symmetries are invariances under 
\bea
\psi_n&\rightarrow& \exp\Big[\im \sum _\A \left(\theta _\A^{(+)}\Gamma^{(+)}_\A+\theta _\A^{(-)}\Gamma^{(-)}_\A\right)\Big]\psi_n\,,\,\,\,\,\,\,\,\, \nonumber\\
\overline{\psi}_n&\rightarrow&\overline{\psi}_n\exp\Big[\im \sum _\A\left(-\theta _\A^{(+)}\Gamma^{(+)}_\A+\theta _\A^{(-)}\Gamma^{(-)}_\A\right)\Big]\,,
\label{sym_naive_all}
\eea
where $\Gamma^{(+)}_\A$ and $\Gamma^{(-)}_\A$ are site-dependent $2\times 2$ matrices:
\begin{eqnarray}
\label{sym_t+}
\Gamma^{(+)}_\A&\in &\left\{\mathbf{1}_2\,,\,\, (-1)^{n_1+ n_2}\gamma_3\,,\,\,(-1)^{\check{n}_\mu}\gamma_\mu \right\}\,,\\
\label{sym_t-}
\Gamma^{(-)}_\A&\in &\left\{(-1)^{n_1 + n_2}\mathbf{1}_2\,,\,\, \gamma_3\,,\,\,(-1)^{n_\mu}\gamma_\mu \right\}\,,
\end{eqnarray}
with $\check{n}_\mu=n_{\nu\neq\mu}$.
The on-site mass term $\bar{\psi}_{n} \psi_{n}$ breaks this $U(2)\times U(2)$ symmetries to the $U(2)$ subgroup, generated by $\Gamma^{(+)}_\A$.
In the presence of the Wilson term the $U(2)\times U(2)$ invariance is broken
to the $U(1)$ invariance under $\mathbf{1}_{2}$ in Eq.~(\ref{sym_t+}).
This generator is vector-type, which means that the Wilson fermion loses all the axial symmetry. 

\begin{figure}
\centering
\includegraphics[height=5cm]{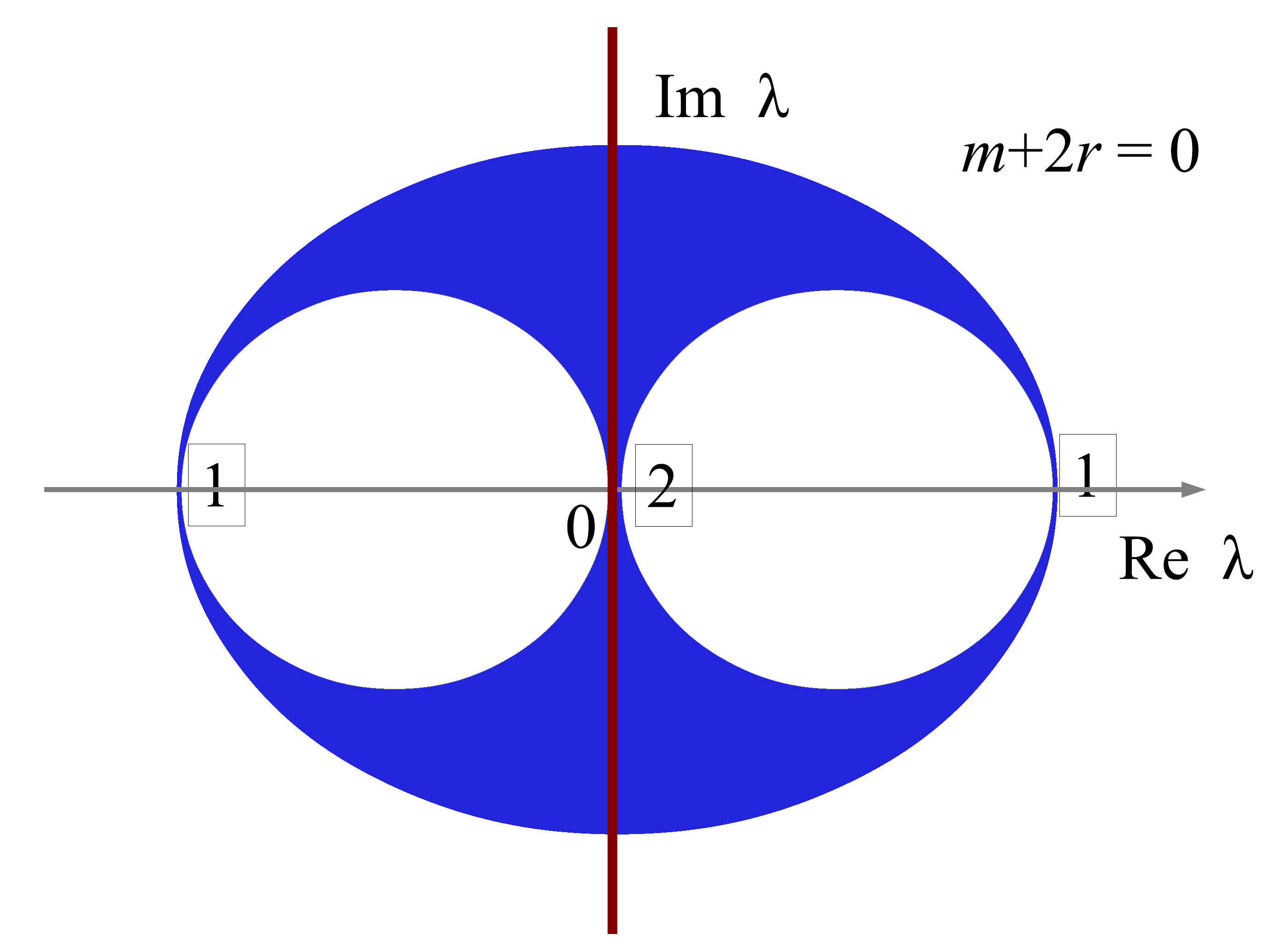} 
\caption{Schematic plot of distribution of the 2d free Wilson Dirac spectrum $\lambda$ with $M_{W}=m+2r=0$ in the complex plane. 
The central branch crosses the origin. The number in each branch stands for
numbers of species at the branch. }
\label{WilD}
\end{figure}

As discussed above, the 2d Wilson term lifts four species into three branches in
the Dirac spectrum in Fig.~\ref{WilD}, and we shall discuss its details in Sec.~\ref{sec:symmetry_spectrum}. 
In Ref.~\cite{Creutz:2011cd, Kimura:2011ik}, it has been shown that the Wilson fermion with the 
condition, 
\be
M_{W}\equiv  m+2r=0,
\ee
has an extra $U(1)$ symmetry besides the usual $U(1)$ vector symmetry.
The Wilson fermion with this condition gives the two-flavor massless fermions, which correspond to the central branch of the Wilson Dirac spectrum as shown in Fig.~\ref{WilD}. 
The fermion lattice action for this case is given by
\begin{equation}
S_{\rm CB}=\sum_{n,\mu}\left(\bar{\psi}_{n}\gamma_{\mu}D_{\mu}\psi_{n}-r\bar{\psi}_{n}C_{\mu}\psi_{n}\right).
\label{CB}
\end{equation}
This action is invariant under the ordinary $U(1)_V$ transformation generated by $\Gamma^{(+)}=\bm{1}_2$, 
\begin{equation}
U(1)_V: \psi_n \mapsto \rme^{\im \alpha}\psi_n,\quad \overline{\psi}_n\mapsto \overline{\psi}_n \rme^{-\im \alpha}, 
\label{eq:ordinary_U1}
\end{equation}
and furthermore there is the extra $U(1)$ symmetry generated by $\Gamma^{(-)}=(-)^{n_1+n_2}\bm{1}_2$, 
\begin{equation}
U(1)_{\overline{V}}: \psi_n\mapsto \rme^{\im (-1)^{n_1+n_2}\beta}\psi_n,\; \overline{\psi}_n\mapsto \overline{\psi}_n \rme^{\im (-1)^{n_1+n_2}\beta}. 
\label{CBsym}
\end{equation}  
The usual Wilson fermion has only the vector symmetry (\ref{eq:ordinary_U1}).
The invariance under (\ref{CBsym})
is restored only with the central-branch condition\footnote{The symmetry associated with $(-1)^{n_1 + n_2}$ is the same as
the chiral symmetry in 2d staggered fermion, which works as the axial rotation. However, we will see in this paper that its role is very different for the central-branch Wilson fermion. } $m+2r=0$.
It is notable that this extra symmetry prohibits the on-site mass term $\bar{\psi}_n\psi_n$, 
and eventually prohibits additive mass renormalization 
as the chiral symmetry in staggered fermion does~\cite{Creutz:2011cd, Kimura:2011ik, Misumi:2012eh}. 
This formulation is regarded as another realization of 
lattice fermions with the remnant of chiral symmetry, 
which means we do not need fine-tuning of the mass parameter.
It is also notable that such symmetry enhancement on the central branch is generic with the flavored-mass fermions~\cite{Creutz:2010bm}.
The other symmetries of this central-branch fermion are common with those of the usual Wilson fermion, including hypercubic symmetry, charge conjugation, parity, time reversal, $\gamma_{3}$-hermiticity and reflection positivity.
Since we will use the lattice translation and rotational symmetry, let us write them down explicitly: The lattice translation, $\mathbb{Z}^2$, is generated by $\psi(x,y)\mapsto \psi(x+1,y)$ and $\psi(x,y)\mapsto \psi(x,y+1)$. The lattice ${\pi\over 2}$ rotation is given by
\be
\psi(x,y)\mapsto \rme^{\im {\pi\over 4}\gamma_3} \psi(y,-x),\; \overline{\psi}(x,y)\mapsto \overline{\psi}(y,-x) \rme^{-\im {\pi\over 4}\gamma_3}. 
\ee
The 4d central-branch fermion is summarized in Appendix.~\ref{4d}, where 4d two-flavor central-branch fermion is also discussed.

\subsection{Symmetry of the Dirac spectrum at central branch}\label{sec:symmetry_spectrum}

Armed with the knowledge about central-branch Wilson fermions, we discuss the symmetry property of the Dirac spectrum, and we shall show that the Dirac determinant is positive definite. 
This shows that the central-branch Wilson fermion is free from the sign problem. 

We assume that our spacetime is set to torus, and approximate it as $\mathbb{Z}^2/(N_x \mathbb{Z}\times N_y\mathbb{Z})$. 
We only consider the case $N_x, N_y$ are even integers, so that $(-)^{x+y}$ is well defined. 
We denote the central-branch Wilson-Dirac operator as 
\be
\mathsf{D}=\sum_{\mu=1,2}(\gamma_\mu D_\mu-r C_\mu). 
\ee
On each site, there is a two-component spinor, so $\mathsf{D}$ is regarded as the linear operator, $\mathsf{D}: \mathbb{C}^{2N_x N_y}\to \mathbb{C}^{2 N_x N_y}$. 
We consider the eigenvalue problem, 
\bea
\mathsf{D} |R_{\lambda}\rangle &=&\lambda |R_{\lambda}\rangle, \\
\langle L_\lambda | \mathsf{D} &=& \lambda \langle L_\lambda|, 
\eea
where $\lambda \in \mathbb{C}$ is called the Dirac eigenvalue, and $|R_\lambda\rangle$ and $\langle L_\lambda|$ are the corresponding right- and left-eigenvectors, respectively. 
For the free theory, we can diagonalize $\mathsf{D}$ by Fourier transformation, and we obtain that 
\be
\lambda(p_x,p_y)=\pm \im \sqrt{\sin^2 p_x+\sin^2 p_y}-r (\cos p_x + \cos p_y),
\ee
where $(p_x,p_y) \bmod 2\pi$ denotes the lattice momentum. Blue shaded region of Fig.~\ref{WilD} shows the distribution of this $\lambda(p_x,p_y)$ in the complex plane. 
We note that $\lambda(p_x,p_y)=0$ only has the two solutions,
\be
(p_x,p_y) = (\pi, 0 ), \; (0,\pi), 
\ee
so there are two gappless fermions at the central branch. 

Let us go back to the discussion for the Dirac operator with gauged link variables. 
As a consequence of $U_{n,\mu}^\dagger=U_{n+\hat{\mu},-\mu}$, we obtain $T_\mu^\dagger=T_{\mu}^{-1}=T_{-\mu}$. 
This ensures the $\gamma_3$-hermiticity of the Wilson-Dirac operator, 
\be
\gamma_3 \mathsf{D} \gamma_3=\mathsf{D}^\dagger. 
\ee
Therefore, by taking the adjoint of the eigenvalue equations, we get 
\bea
\mathsf{D} \gamma_3|L_{\lambda}\rangle &=&\lambda^* \gamma_3 |L_{\lambda}\rangle, \\
\langle R_\lambda |\gamma_3 \mathsf{D} &=& \lambda^* \langle R_\lambda|\gamma_3, 
\eea
This shows that when $\lambda \in \mathbb{C}\setminus \mathbb{R}$ is in the Dirac spectrum so is $\lambda^*$.

So far, we have seen a generic feature of any Wilson fermion by $\gamma_3$-hermiticity. 
The existence of the site-dependent $U(1)$ symmetry, $U(1)_{\overline{V}}$, is the special feature of the central-branch Wilson fermion. 
This means that the central-branch Wilson-Dirac operator satisfies
\be
\mathsf{D} (-)^{x+y}=- (-)^{x+y} \mathsf{D}. 
\ee
Using this anti-commutation relation, we obtain 
\bea
\mathsf{D} (-)^{x+y}|R_{\lambda}\rangle &=&-\lambda (-)^{x+y}|R_{\lambda}\rangle, \\
\langle L_\lambda | (-)^{x+y} \mathsf{D} &=& -\lambda \langle L_\lambda|(-)^{x+y}. 
\eea
This shows that if $\lambda \in \mathbb{C}\setminus\{0\}$ is in the Dirac spectrum so is $-\lambda$. 
These symmetries explain why Fig.~\ref{WilD} is symmetric under $\mathrm{Re}(\lambda)\mapsto -\mathrm{Re}(\lambda)$ and $\mathrm{Im}(\lambda)\mapsto -\mathrm{Im}(\lambda)$. 

Now, we would like to show that the central-branch Wilson fermion has no sign problem, i.e. 
\be
\mathrm{det}(\mathsf{D})\ge 0. 
\ee 
We emphasize that this is an important property of this fermion when we consider the Monte Carlo simulation of lattice gauge theory. 
In order to prove this, it is useful to introduce the hermitian Wilson-Dirac operator, 
\be
H=\gamma_3 \mathsf{D}. 
\ee
The $\gamma_3$-hermiticity of $\mathsf{D}$ ensures that $H^\dagger =H$, so its spectrum is in real values. 
The $U(1)_{\overline{V}}$ symmetry gives $H(-)^{x+y}=-(-)^{x+y}H$, so the non-zero spectrum forms the pair with the opposite sign. 
When there are no zero eigenvalues, we can label the spectrum as 
\be
\{\pm \ve_i\}_{i=1,\ldots,N_x N_y}. 
\ee
Since $N_x N_y$ is an even integer, we obtain that 
\be
\mathrm{det}(\mathsf{D})=\mathrm{det}(H)=\prod_{i=1}^{N_x N_y}\ve_i (-\ve_i) =(-1)^{N_x N_y}\prod_{i=1}^{N_xN_y} \ve_i ^2>0. 
\ee
If there are some zero eigenvalues, $\mathrm{det}(\mathsf{D})=0$. We have shown the semi-positivity of $\mathrm{det}(\mathsf{D})$. 

We note that the same argument can be used for $4$d central-branch Wilson fermion, too, and the Dirac determinant is again positive semi-definite. 

\section{Analytical study of low-energy effective theory}\label{sec:anomaly}

In this section, we study the property of low-energy effective theory of the lattice Schwinger model with the central-branch Wilson fermion. 
By using the low-energy approximation, we make the connection between the lattice gauge theory and the continuum field theory. 
Using this approximation, we can translate the exact symmetry on lattice into the emergent internal symmetry on continuum, and we compute the 't~Hooft anomaly of the symmetry.

\subsection{Low-energy approximation}

In order to find the symmetry structure of this lattice fermion, let us make connection with the continuum description. 
Let us write down the explicit form of the central-branch Wilson fermion (\ref{CB}), 
\be
S_{\mathrm{CB}}={1\over 2}\sum_{n}\sum_{\mu=1,2}(\overline{\psi}_{n}\gamma_\mu \psi_{n+\hat{\mu}}-\overline{\psi}_{n+\hat{\mu}}\gamma_\mu \psi_n)-{r\over2} \sum_n \sum_\mu \overline{\psi}_n(\psi_{n+\hat{\mu}}+\psi_{n-\hat{\mu}}), 
\ee
and, importantly, the on-site term $\overline{\psi}_n\psi_n$ is gone thanks to the condition $m+2r=0$. 

Since two lattice momenta, $(p_x,p_y)=(\pi,0)$ and $(0,\pi)$, give gappless modes, we can expect that the following scale separation works nicely:
\bea
\psi(x,y)&=:&(-1)^x\,\psi_1(x,y)+(-1)^y\, \psi_2(x,y), \label{eq:lowE_approx_1}\\
\overline{\psi}(x,y)&=:&(-1)^x\, \overline{\psi}_1(x,y)-(-1)^y\, \overline{\psi}_2(x,y).  \label{eq:lowE_approx_2} 
\eea
$\psi_{1,2}$ only contains the low-momentum modes, and the staggered phases in front describe the fast (i.e. lattice-scale) oscillations. 
In (\ref{eq:lowE_approx_2}), we multiply the extra $(-1)$ factor in front of $\overline{\psi}_2$, and the reason will become evident in the following computation of the low-energy effective action. 
We call this label, $1,2$, as the flavor label, and each flavor is the two-component spinor field. We denote 
\be
\Psi= \begin{pmatrix}
\psi_1\\
\psi_2
\end{pmatrix},\quad 
\overline{\Psi}=\begin{pmatrix}
\overline{\psi}_1 & \overline{\psi}_2
\end{pmatrix}. 
\ee

Let us compute the form of the effective action by substituting (\ref{eq:lowE_approx_1}) and (\ref{eq:lowE_approx_2}). The naive kinetic term gives
\bea
&&{1\over 2}\sum_{n}\sum_{\mu=1,2}(\overline{\psi}_{n}\gamma_\mu \psi_{n+\hat{\mu}}-\overline{\psi}_{n+\hat{\mu}}\gamma_\mu \psi_n)\nonumber\\
&=&\sum_{(x,y)} \sum_{i=1,2}\overline{\psi}_i [-\gamma_1 \p_x+\gamma_2 \p_y] \psi_i(x,y)\nonumber\\
&&+\sum_{(x,y)} (-)^{x+y} [\overline{\psi}_1 \gamma_1\p_x \psi_2+\overline{\psi}_2 \gamma_1\p_x \psi_1-\overline{\psi}_1\gamma_2\p_y \psi_2-\overline{\psi}_2\gamma_2\p_y \psi_1]. 
\eea
Because of the staggered phases in (\ref{eq:lowE_approx_1}) and (\ref{eq:lowE_approx_2}), we get the $(-1)$ phase for the $x$-derivative acting on $\psi_1$, and for the $y$-derivative acting on $\psi_2$. 
This means that the kinetic term of $\psi_1$ and $\psi_2$ has the opposite $(-1)$ sign, so we put the extra $(-1)$ sign in front of $\overline{\psi}_2$ in (\ref{eq:lowE_approx_2}) in order to eliminate it. 
Moreover, we redefine the gamma matrices as 
\be
\gamma_1^{\mathrm{(new)}}=-\gamma_1,\;\; \gamma_2^{(\mathrm{new})}=\gamma_2, 
\ee
so that we obtain the usual kinetic term, $\slashed{\p}=\gamma^{(\mathrm{new})}_{\mu}\p_\mu$, in the first line. 
Because of the staggering phase, we expect that the second term should become negligible in the continuum limit. 
We still keep this term in our analysis because this term is important in order to understand which symmetry is genuine at the lattice level. 

The Wilson + mass term gives
\be
-{r\over2} \sum_n \sum_\mu \overline{\psi}_n(\psi_{n+\hat{\mu}}+\psi_{n-\hat{\mu}})=-{r\over 2}\sum_{(x,y)}\left[\overline{\Psi} (\p_y^2-\p_x^2)\Psi-(-)^{x+y}\overline{\Psi}\tau_1 (\p_y^2-\p_x^2)\Psi\right]. 
\ee
Combining them, we get 
\be
S= \sum_{(x,y)}\left\{ \overline{\Psi} \left[\slashed{\p}-{r\over 2}(\p_y^2-\p_x^2)\right] \Psi-(-)^{x+y} \overline{\Psi}\tau_1 \left[\slashed{\p}-{r\over 2}(\p_y^2-\p_x^2)\right]\Psi\right\}. 
\label{eq:action_EFT}
\ee
Here, $\tau_i$ is the Pauli matrices in the flavor space.

\subsection{From lattice symmetry to internal symmetry}

We have obtained the connection between lattice fermion and the continuum effective description. This allows us to identify the role of lattice symmetry in the continuum. 
We shall see that both lattice translation and lattice ${\pi\over 2}$ rotation induce important internal symmetries in the continuum limit. 

\subsubsection{Vector-like symmetry}

First, let us discuss the vector-like symmetry in the continuum limit. 
We claim that the vector-like symmetry of this system is given by 
\be
{U(1)_V \times [U(1)_{\overline{V}}\rtimes (\mathbb{Z}_2)_{\mathrm{lat.\,trans.}}] \over (\mathbb{Z}_2)_F}. 
\ee
Here, $(\mathbb{Z}_2)_F$ is generated by the fermion parity transformation, $\Psi\mapsto -\Psi$.
We will gauge $U(1)_V$ in the lattice Schwinger model, so the vector-like global symmetry is given by 
\be
{U(1)_{\overline{V}}\rtimes (\mathbb{Z}_2)_{\mathrm{lat.\,trans.}} \over (\mathbb{Z}_2)_F}=O(2) \subset {SU(2)_{\mathrm{flavor}} \over (\mathbb{Z}_2)_F}, 
\ee
and this is the subgroup of the flavor $SU(2)/\mathbb{Z}_2(= SO(3))$ symmetry.

\noindent
\textbf{Lattice on-site symmetry:} 
We first discuss the lattice on-site symmetry. 
As we have seen in the previous section, there are two $U(1)$ symmetries, $U(1)_{V}$ and $U(1)_{\overline{V}}$. 
The $U(1)_V$ symmetry~(\ref{eq:ordinary_U1}) gives the flavor-independent $U(1)$ symmetry,
\be
\Psi \mapsto \rme^{\im \alpha} \Psi,\; \overline{\Psi}\mapsto \overline{\Psi}\rme^{-\im \alpha}. 
\ee
The site-dependent $U(1)$ symmetry~(\ref{CBsym}), $U(1)_{\overline{V}}$, gives the flavored $U(1)$ symmetry,  
\be
\Psi \mapsto \rme^{\im \beta \tau_1}\Psi,\; \overline{\Psi}\mapsto \overline{\Psi}\rme^{-\im \beta \tau_1}. 
\label{eq:extraU1_continuum}
\ee
Since this result is less trivial compared with that of $U(1)_V$, let us show it explicitly. 
The site-dependent $U(1)$ symmetry (\ref{CBsym}) is given by $\psi\mapsto\rme^{\im (-)^{x+y}\beta}\psi$, $\overline{\psi}\mapsto \overline{\psi}\rme^{\im (-)^{x+y}\beta}$, and then the infinitesimal transformation is 
\bea
\delta \psi&=& \im (-)^{x+y} ((-)^x \psi_1+(-)^y \psi_2)\nonumber\\
&=& \im  ((-)^x \psi_2+(-)^y\psi_1). 
\eea
Therefore, we find that 
\be
\delta \Psi = \begin{pmatrix}
\delta \psi_1\\
\delta \psi_2
\end{pmatrix} =\begin{pmatrix}
\im \psi_2\\
\im \psi_1
\end{pmatrix} = \im \tau_1 \Psi. 
\ee
The transformation for the conjugate fields can be obtained in the similar way, but we should notice that there is an extra $(-1)$ sign in front of $\overline{\psi}_2$ in (\ref{eq:lowE_approx_2}): 
\bea
\delta \overline{\psi}&=& ((-)^x \overline{\psi}_1- (-)^y \overline{\psi}_2) (\im (-)^{x+y})\nonumber\\
&=& (-)^x (-\overline{\psi}_2) - (-)^y (-\overline{\psi}_1). 
\eea
As a result, $\overline{\Psi}$ obeys the conjugate representation, $\delta \overline{\Psi}=-\im \overline{\Psi}\tau_1$. 
Since both $\pi$ rotations of $U(1)_V$ and $U(1)_{\overline{V}}$ define the fermion parity, $(\mathbb{Z}_2)_F$, the symmetry with faithful action becomes the quotient group. 

We can easily understand the existence of this extra symmetry by looking at the second term of (\ref{eq:action_EFT}), e.g., $(-1)^{x+y}\overline{\Psi}\tau_1 \slashed{\p}\Psi$. 
We naively expect $SU(2)$ symmetry for two degenerate Dirac fermions, but those two flavors are connected at the lattice energy scale. The continuous rotations along $\tau_2$ and $\tau_3$ are explicitly broken because of this effect. 
Since this term is oscillatory at the lattice scale, we expect that it becomes negligible in the continuum limit. This point requires the further study with numerical simulations. 

\noindent
\textbf{Lattice translation:} 
Interestingly, this is not the end of the story. The lattice translation induces a discrete flavored rotation along $\tau_3$, and we will call it $(\mathbb{Z}_2)_{\mathrm{lat.\,trans.}}$ since this is $\mathbb{Z}_2$ symmetry emerging from the lattice translation. 

The lattice translation by one unit along $y$ direction gives 
\bea
\psi(x,y+1)&=&(-)^x \psi_1(x,y+1)+(-)^{y+1}\psi_2(x,y+1) \nonumber\\ 
&\simeq& (-)^x \psi_1(x,y)-(-)^y \psi_2(x,y). 
\eea
In the second approximate equality, we neglect the difference of $y$ by one lattice unit inside $\psi_{1,2}$, because $\psi_{1,2}$ describes the low-momentum modes. 
Therefore, in the low-momentum limit, the lattice translation $\mathbb{Z}\times \mathbb{Z}$ enhances to $\mathbb{Z}_2\times \mathbb{R}^2$. The $\mathbb{Z}_2$ internal transformation is given as 
\be
(\mathbb{Z}_2)_{\mathrm{lat.\,trans.}}: \Psi \mapsto \tau_3 \Psi,\; \overline{\Psi}\mapsto \overline{\Psi}\tau_3. 
\ee
We note that this is not realized as an on-site lattice symmetry. Since $\tau_3\tau_1\tau_3=-\tau_1$, this transformation flips the sign of the second term of (\ref{eq:action_EFT}). To make the action invariant, we also need to flip the staggering phase $(-)^{x+y}$. Therefore, it should be combined with the lattice translation by one unit. 

\subsubsection{Chiral symmetry}

Next, we consider about the chiral symmetry. We show that the lattice ${\pi\over 2}$ rotation induces $\mathbb{Z}_2$ chiral transformation, $\Psi\mapsto \gamma_3 \Psi$ and $\overline{\Psi}\mapsto \overline{\Psi}(-\gamma_3)$. 
We would like to emphasize that this is the best \textit{flavor-singlet} chiral symmetry on the lattice for this system. 
Since we have the on-site $U(1)_V$ symmetry, we have no obstruction to gauge it. Since there are two massless Dirac fermions $\Psi$, the axial $U(1)$ symmetry is broken to the above $\mathbb{Z}_2$ symmetry by Adler-Bell-Jackiw anomaly~\cite{Adler:1969gk, Bell:1969ts}, so our lattice model has the same flavor-singlet chiral symmetry with that of continuum description of the two-flavor Schwinger model\footnote{In the continuum, we can consider the $U(1)$ gauge theory only with charge-$q$ Dirac fermions, which leads to the larger discrete chiral symmetry, $\mathbb{Z}_{2q}$~\cite{Anber:2018jdf,   Anber:2018xek, Armoni:2018bga, Misumi:2019dwq} (see also~\cite{Pantev:2005zs, Pantev:2005wj}). However, due to the fact that we have an on-site $U(1)$ symmetry that can be gauged as charge-$1$ theory, we can only have $\mathbb{Z}_2$ chiral symmetry in our lattice construction when we consider no fine-tuning. }. 
 
Now, let us make a connection with lattice rotational symmetry and discrete chiral symmetry. 
The lattice ${\pi\over 2}$ rotation is given by
\be
\psi(x,y)\mapsto \rme^{\im {\pi\over 4}\gamma_3} \psi(y,-x),\; \overline{\psi}(x,y)\mapsto \overline{\psi}(y,-x) \rme^{-\im {\pi\over 4}\gamma_3}. 
\ee
Substituting this into the low-energy expression, we find that 
\be
\Psi(x,y)\mapsto \rme^{\im {\pi\over 4} (-\gamma_3^{(\mathrm{new})})}\otimes \tau_1 \Psi(y,-x),\; 
\overline{\Psi}(x,y)\mapsto  \overline{\Psi}(y,-x) \rme^{-\im {\pi\over 4} (-\gamma_3^{(\mathrm{new})})}\otimes (-\tau_1). 
\ee
Appearance of $\tau_1$ is due to the exchange of staggering phases $(-)^x\leftrightarrow (-)^y$, and this is not so important. 
We also use the fact that the gamma matrices are redefined as $\gamma_\mu^{(\mathrm{new})}=(-)^\mu \gamma_\mu$ in taking the continuum limit at the central branch. 
This flips the sign of the spin rotation, $\Sigma={1\over 4}[\gamma_1,\gamma_2]={\im\over 2}\gamma_3$, in the exponent. 
As a result, the spin rotation direction becomes opposite. 
To resolve this issue, we note that 
\be
\rme^{\im {\pi\over 4}(-\gamma_3^{(\mathrm{new})})}={1-\im \gamma_3^{(\mathrm{new})}\over \sqrt{2}}=-\im \gamma_3^{(\mathrm{new})}{1+\im \gamma_3^{(\mathrm{new})}\over \sqrt{2}}=-\im \gamma_3^{(\mathrm{new})}\rme^{\im{\pi\over 4} \gamma_3^{(\mathrm{new})}}. 
\ee
Using this, the lattice ${\pi\over 2}$ rotation turns out to be a combination of the spacetime ${\pi\over 2}$ rotation and the discrete chiral transformation in the continuum limit:
\bea
\Psi(x,y)&\mapsto& \gamma_3^{(\mathrm{new})}\rme^{\im{\pi\over 4} \gamma_3^{(\mathrm{new})}}\Psi(y,-x),\; \nonumber\\
\overline{\Psi}(x,y)&\mapsto& \overline{\Psi}(y,-x)\rme^{-\im{\pi\over 4} \gamma_3^{(\mathrm{new})}} (-\gamma_3^{(\mathrm{new})}). 
\eea
Here, we eliminate $\im \tau_1={\rme^{\im {\pi\over 2} \tau_1 }}$ by a site-dependent $U(1)$ transformation, and this simplifies the expression. 
We can understand this invariance as follows. 
Look at the effective action~(\ref{eq:action_EFT}), then the kinetic term with the first-order derivative $\slashed{\p}$ has the invariance under axial symmetry. However, the kinetic terms with the second-order derivative, $\p_y^2-\p_x^2$, break the axial symmetry completely. 
We also notice that it also breaks naive ${\pi\over 2}$ rotation, since $\p_y^2-\p_x^2$ flips its sign. The idea is that the combination of them is the symmetry. 

Assuming that $\mathbb{Z}_8$ lattice rotation\footnote{This is $\mathbb{Z}_8$ instead of $\mathbb{Z}_4$ because $2\pi$ rotation of fermions gives $(-1)^F$, and we get $Spin(2)$ for the Lorentz symmetry for fermions in continuum instead of $SO(2)$. } enhances to the Lorentz symmetry in the continuum limit, this indicates that the discrete chiral transformation also emerges:
\be
(\mathbb{Z}_8)_{\mathrm{lattice\; rot.}}\xrightarrow{\mathrm{enhance}} (\mathbb{Z}_2)_\chi \times Spin(2). 
\ee
As a result, we have the following internal symmetry in the continuum description for the central-branch Wilson fermion, 
\be
G_{\mathrm{CB\; fermion}}={U(1)_V \times [U(1)_{\overline{V}}\rtimes (\mathbb{Z}_2)_{\mathrm{lat.\;trans.}}]\over (\mathbb{Z}_2)_F}\times (\mathbb{Z}_2)_\chi, 
\ee
and this originates from the exact lattice symmetry. 
We gauge the $U(1)_V$ symmetry by introducing the link variables, and then the global symmetry is divided by $U(1)$ and becomes
\be
G=G_{\mathrm{CB\; fermion}}/U(1)_V={U(1)_{\overline{V}}\rtimes (\mathbb{Z}_2)_{\mathrm{lat.\;trans.}}\over (\mathbb{Z}_2)_F}\times (\mathbb{Z}_2)_\chi. 
\ee

\subsection{Flavor singlet and non-singlet mass terms}

We will show that the symmetry $G$ has the 't~Hooft anomaly and gives an important constraint on non-perturbative low-energy physics. 
Especially, its existence prohibits to create the mass gap without having degenerate ground states. 
This condition would be  obviously violated if we could write down the fermion bilinear mass terms, because we can obtain the single gapped ground state by sending such mass parameters to infinite. 
As a corollary, we cannot write down the mass term that is invariant under $G$. Since we can find this conclusion in a more elementary way than computing 't~Hooft anomaly, let us give a detailed discussion about it in this section.

The fermion bilinear operator with $U(1)_V$ symmetry has the form\footnote{When we dynamically gauge $U(1)_{\overline{V}}$,  we have to insert the Wilson lines for gauge invariance. Just for notational simplicity, we consider the free fermion case. }, 
\be
\overline{\psi}(x+n_1,y+n_2)\gamma_i\psi(x,y).
\ee 
In order to have the $U(1)_{\overline{V}}$ symmetry, we must set $n_1+n_2$ an odd integer. 
The lattice translational symmetry forbids to multiply the staggering phases, such as $(-)^{x,y}$. 
We further can use the lattice ${\pi\over 2}$ rotation to constrain the possible terms. 
For example, if $\gamma_i=1$, these constraints require that it should appear in the combination,
\begin{align}
\left[\overline{\psi}(x+n_1,y+n_2)+\overline{\psi}(x+n_2,y-n_1)+\overline{\psi}(x-n_1,y-n_2)+\overline{\psi}(x-n_2,y+n_1)\right]\psi(x,y). 
\end{align}
Substituting the low-energy expression (\ref{eq:lowE_approx_1}) and (\ref{eq:lowE_approx_2}), we obtain the leading term as 
\be
2((-1)^{n_1}+(-1)^{n_2})\left(\overline{\psi}_1 \psi_1-\overline{\psi}_2\psi_2\right). 
\ee
Since $n_1+n_2$ has to be an odd integer, this leading term cancels as $(-1)^{n_1}+(-1)^{n_2}=0$, and it starts from the second-order derivatives in the low-energy limit. 
For other gamma matrices, it is straightforward to check that the leading term also starts from the derivatives, so we cannot obtain the mass term that is invariant under all the symmetries. 
This argument shows that the symmetry $G$ prohibits any type of fermion bilinear mass terms. 

In the previous studies~\cite{Creutz:2011cd, Kimura:2011ik, Misumi:2012eh}, it was shown that the on-site mass term, such as $\overline{\psi}_n \psi_n$, is prohibited by $U(1)_{\overline{V}}$. 
Since this on-site mass is clearly invariant under the lattice translation and rotation, this is consistent with the above discussion. 
We would like to emphasize that the above discussion discloses importance of lattice symmetries and generalizes the results in those previous studies, because $U(1)_{\overline{V}}$ itself allows us to add the off-site mass term.

In order to introduce the fermion bilinear mass, we have to break at least one of $U(1)_{\overline{V}}$, lattice translation, and lattice rotational symmetries. Below, let us explicitly show how the mass terms break these symmetry. 

\noindent
\textbf{Flavor-singlet mass:} 
We first consider the flavor-singlet mass, 
\be
m_{0} \overline{\Psi} \Psi. 
\ee
This breaks the discrete chiral symmetry $(\mathbb{Z}_2)_\chi$, while the vector-like symmetry is kept intact. 
In the lattice description, this means that we break the lattice ${\pi\over 2}$ rotation explicitly, and other symmetries are unbroken. 
We have such a term as 
\be
-{m_0\over 4}\sum_{(x,y)} \left[\overline{\psi}(x+1,y)+\overline{\psi}(x-1,y)-\overline{\psi}(x,y+1)-\overline{\psi}(x,y-1)\right]\psi(x,y). 
\ee
Indeed, the site-dependent $U(1)$ and lattice translational symmetries are still exact for this perturbation, and the substitution of (\ref{eq:lowE_approx_1}) and (\ref{eq:lowE_approx_2}) gives the flavor-singlet mass term.

\noindent
\textbf{Flavor non-singlet mass:}
We can consider a mass term that does not break discrete chiral symmetry. It is the flavor non-singlet mass term, 
\be
m_3 \overline{\Psi}\tau_3 \Psi. 
\ee
This breaks both $U(1)_{\overline{V}}$ and $(\mathbb{Z}_2)_{\chi}$ separately, but their combination, 
\be
\Psi\mapsto \gamma_3\otimes (\im \tau_1)\Psi,\quad \overline{\Psi}\mapsto \overline{\Psi} (-\gamma_3)\otimes (-\im \tau_1),
\ee 
is invariant. This mass term is given by 
\be
m_3 \overline{\psi}(x,y)\psi(x,y)\simeq m_3 \left(\overline{\psi}_1\psi_1-\overline{\psi}_2\psi_2\right),
\ee
and this breaks the site-dependent $U(1)$ symmetry, $U(1)_{\overline{V}}$.  
We note that the flavored mass, introduced in Refs.~\cite{Creutz:2010bm,Misumi:2012eh,M1}, also does the same role. For example, the low-energy approximation of the two-link tensor mass, $M_T=C_1C_2$, breaks the $U(1)_{\overline{V}}$ symmetry in the same manner. 

Since we do not have the full $SU(2)$ flavor rotation as the symmetry, we need to consider separately for another flavored mass term,
\be
m_1 \overline{\Psi}\tau_1 \Psi. 
\ee
This keep $U(1)_{\overline{V}}$ intact, but breaks the others into the diagonal subgroup, $(\mathbb{Z}_2)_{\mathrm{lat.\;trans.}}\times (\mathbb{Z}_2)_\chi \to \mathbb{Z}_2$. 
We indeed obtain this by low-energy approximation of 
\be
-{m_1\over 4}\sum_{(x,y)} (-)^{x+y}\left[\overline{\psi}(x+1,y)+\overline{\psi}(x-1,y)-\overline{\psi}(x,y+1)-\overline{\psi}(x,y-1)\right]\psi(x,y). 
\ee
This breaks lattice translation and rotation separately, but some combinations are still unbroken. 

\subsection{Anomaly matching and low-energy physics}
\label{sec:anomaly_matching}

We have seen that we cannot write down the fermion mass term without violating the symmetry $G$. 
Using 't~Hooft anomaly matching condition, we can show the more strong fact that the system cannot have the gapped unique ground state. 
An 't~Hooft anomaly of symmetry $G$ is defined as follows: We define the partition function with the background $G$-gauge field $A$, $Z[A]$, and consider the $G$-gauge transformation, $A\mapsto A+\diff \theta$. If the phase of the partition function behaves as 
\be
Z[A+\diff \theta]=Z[A]\exp(\im \mathcal{A}[A,\theta]), 
\ee
this anomalous phase $\mathcal{A}$ is called the 't~Hooft anomaly\footnote{For clarity, we note that the symmetry is not explicitly broken even if the 't~Hooft anomaly is present. }. Importantly, 't~Hooft anomaly is invariant under the renormalization-group flow, and thus non-trivial 't~Hooft anomaly requires the non-trivial infrared dynamics, such as gapless excitations, spontaneous symmetry breaking, or intrinsic topological order.

In this section, following Refs.~\cite{Komargodski:2017dmc, Komargodski:2017smk, Sulejmanpasic:2018upi, Yao:2018kel, Tanizaki:2018xto}, we first show that the symmetry $G$ of the 2d central-branch Wilson fermion has $\mathbb{Z}_2$ 't~Hooft anomaly in continuum. 
This anomaly is the field theoretic realization of the LSM theorem for spin-$1/2$ chain, and, using this fact, we discuss the possible low-energy behaviors in comparison with the exact solution of Heisenberg $XYZ$ model. 
We also discuss its connection to the Aoki phase~\cite{Aoki:1983qi, Aoki:1986xr, Aoki:1987us} of 2d lattice Gross-Neveu model with the Wilson fermion. 

\subsubsection{'t~Hooft anomaly and comparison with Heisenberg $XYZ $ model}
\label{sec:tHooft_anomaly}

We consider the $U(1)$ gauge theory with the central-branch Wilson fermion, and then the system has the global symmetry $G$. Let us introduce the background gauge field for the vector-like symmetry, and we will see that $(\mathbb{Z}_2)_\chi$ is anomalously broken~\cite{Komargodski:2017dmc, Komargodski:2017smk, Sulejmanpasic:2018upi, Yao:2018kel, Tanizaki:2018xto}. 

As the vector-like symmetry, we especially pay attention to the subgroup, 
\be
\mathbb{Z}_2\times \mathbb{Z}_2\simeq {(\mathbb{Z}_4)_{\overline{V}}\rtimes (\mathbb{Z}_2)_{\mathrm{lat.\;trans.}}\over (\mathbb{Z}_2)_F} \subset {U(1)_{\overline{V}}\rtimes (\mathbb{Z}_2)_{\mathrm{lat.\; trans.}}\over (\mathbb{Z}_2)_F}. 
\ee
The background gauge field can be realized as the twisted boundary condition on the two-torus $T^2$. 
We twist the fermion boundary conditions by $\im \tau_1\in (\mathbb{Z}_4)_{\overline{V}}$ along $x$-direction, and by $\tau_3\in (\mathbb{Z}_2)_{\mathrm{lat.\;trans.}}$ along $y$-direction:
\bea
\Psi(x+L, y)&=&\rme^{\im \alpha_1(y)} (\im \tau_1) \Psi(x,y), \\
\Psi(x,y+L)&=& \rme^{\im \alpha_2(x)}\tau_3 \Psi(x,y). 
\eea
Here, $\alpha_1$ and $\alpha_2$ denote the transition functions of $U(1)_V$ gauge symmetry along $x$ and $y$ directions, respectively, and they are $2\pi$-periodic scalars. 
These transition functions are introduced because the fermion wave function needs to be periodic only up to the $U(1)$ gauge transformations. 
We can relate $\Psi(x+L,y+L)$ and $\Psi(x,y)$ in two ways: 
\bea
\Psi(x+L,y+L)&=& \rme^{\im \alpha_1(y+L)}(\im \tau_1)\Psi(x,y+L)\nonumber\\
&=& \rme^{\im (\alpha_1(y+L)+\alpha_2(x))} (\im \tau_1\tau_3)\Psi(x,y),\\
\Psi(x+L,y+L)&=& \rme^{\im \alpha_2(x+L)}\tau_3 \Psi(x+L,y)\nonumber\\
&=& \rme^{\im (\alpha_1(y)+\alpha_2(x+L))} (\im \tau_3\tau_1)\Psi(x,y). 
\eea
We note that $\tau_1 \tau_3=-\tau_3\tau_1$. For consistency, the transition functions must satisfy
\be
\alpha_1(y+L)+\alpha_2(x)=\alpha_1(y)+\alpha_2(x+L)+\pi\quad \bmod 2\pi. 
\ee
We can represent the difference of gauge fields, $a=a_x(x,y)\diff x+a_y(x,y)\diff y$, between $x=L$ and $x=0$, etc., using the transition functions as 
\bea
a(x=L,y)-a(x=0,y)&=&\p_y \alpha_1(y) \diff y,\nonumber\\ a(x,y=L)-a(x,y=0)&=&\p_x \alpha_2(x)\diff x. 
\eea
Therefore, under this twisted boundary condition, the topological charge is fractionalized: 
\bea
{1\over 2\pi}\int_{T^2}\diff a &=& {1\over 2\pi}\left(\int_0^L \diff y (a_y(L,y)-a_y(0,y))-\int_0^L\diff x(a_x(x,L)-a_x(x,0))\right)\nonumber\\
&=&{1\over 2\pi}(\alpha_1(L)-\alpha_1(0)-\alpha_2(L)+\alpha_2(0))\nonumber\\
&\in& {1\over 2}+\mathbb{Z}. 
\eea
Since the system has two-flavor Dirac fermion, the index theorem tells us that there is an odd number of zero modes\footnote{We note that these odd number of fermionic zero modes appear due to the twisted boundary condition. This twisted boundary condition is introduced in order to detect the 't~Hooft anomaly. When performing the numerical simulation of this system, we can use the periodic boundary condition, and then the semi-positivity of the Dirac determinant holds. }. As a result, the partition function with the twisted boundary condition flips its sign under the discrete chiral transformation, 
\be
(\mathbb{Z}_2)_\chi: \calZ_{\mathrm{twisted}}\mapsto -\mathcal{Z}_{\mathrm{twisted}}, 
\ee
which is nothing but the mixed 't~Hooft anomaly. 
This anomaly is the field-theoretic realization of the LSM theorem \cite{Lieb:1961fr, Affleck:1986pq}.

Let us discuss the possible low-energy physics by requiring the anomaly matching condition. 
In $(1+1)$ dimensions, there are two ways to match this anomaly:
\begin{itemize}
\item gapless excitations, or
\item two vacua by spontaneous breaking of discrete symmetry. 
\end{itemize}
In order to get some insight about the possible low-energy behavior, we summarize the result of the Heisenberg $XYZ$ spin-$1/2$ chain (for details, see the textbook, e.g.,~\cite{takahashi1999thermodynamics}):
\be
\hat{H}=-\sum_\ell (J_x \hat{X}_\ell \hat{X}_{\ell+1}+ J_y \hat{Y}_\ell \hat{Y}_{\ell+1}+J_z \hat{Z}_\ell \hat{Z}_{\ell+1}). 
\ee
$J_{x,y,z}$ denote the coupling constants, and $\hat{X}_\ell, \hat{Y}_\ell, \hat{Z}_\ell$ are Pauli matrices for the spin at site $\ell$. 
Generically, this model has the on-site spin symmetry, $(\mathbb{Z}_2\times \mathbb{Z}_2)_{\mathrm{spin}}$, and it has $\mathbb{Z}_2$ mixed anomaly with lattice translation. 
We can summarize the correspondence between symmetries of our lattice $U(1)$ gauge theory and those of the Heisenberg chain as follows:
\be
\begin{tabular}{ c | c | c  } 
Low-energy description &  $XYZ$ spin chain & Our lattice formulation
 \\\hline 
$SO(3)_V$ or its $O(2)$ subgroup & $(\mathbb{Z}_2\times \mathbb{Z}_2)_{\mathrm{spin}}$ & $U(1)_{\overline{V}}\rtimes (\mathrm{lattice\;trans.})$ \\
$(\mathbb{Z}_2)_\chi$ &  lattice translation & lattice rotation
\end{tabular}
\label{eq:correspondence}
\ee
When $J_i\not=J_j$ for $i\not=j$, the system has two ground states and the anomaly is matched by discrete symmetry breaking. It depends on the couplings $J_{x,y,z}$ whether the anomaly is matched by breaking $(\mathbb{Z}_2\times \mathbb{Z}_2)_{\mathrm{spin}}$ (ferromagnetic phase) or by breaking lattice translation (anti-ferromagnetic phase). 
When $J\equiv J_x=J_y\not=J_z$, the model is called the $XXZ$ spin chain and has an enlarged spin symmetry, $SO(2)\rtimes \mathbb{Z}_2$. This enlarged $SO(2)$ corresponds to $U(1)_{\overline{V}}$ in our lattice model, and these two models have exactly the same symmetry structure by the above correspondence. If $|J_z/J|<1$, the system is in the gappless phase with spinon, spin-wave and bound-state excitations, while if $|J_z/J|>1$ the anomaly is matched by two vacua due to discrete symmetry breaking. 

Our anomaly matching argument shows that the lattice Schwinger model with the central-branch fermion belongs to the same universality class, and we do not need fine-tuning of bare parameters. 

\subsubsection{Aoki phase of 2d lattice Gross-Neveu model with Wilson fermion}

It would be useful to compare our result with the preceding studies on the phase structure of Wilson fermion. 
Such studies are very important to understand if the lattice regularized theory has the correct continuum limit when we perform the numerical Monte Carlo simulation. 

\begin{figure}
\centering
\includegraphics[height=6cm]{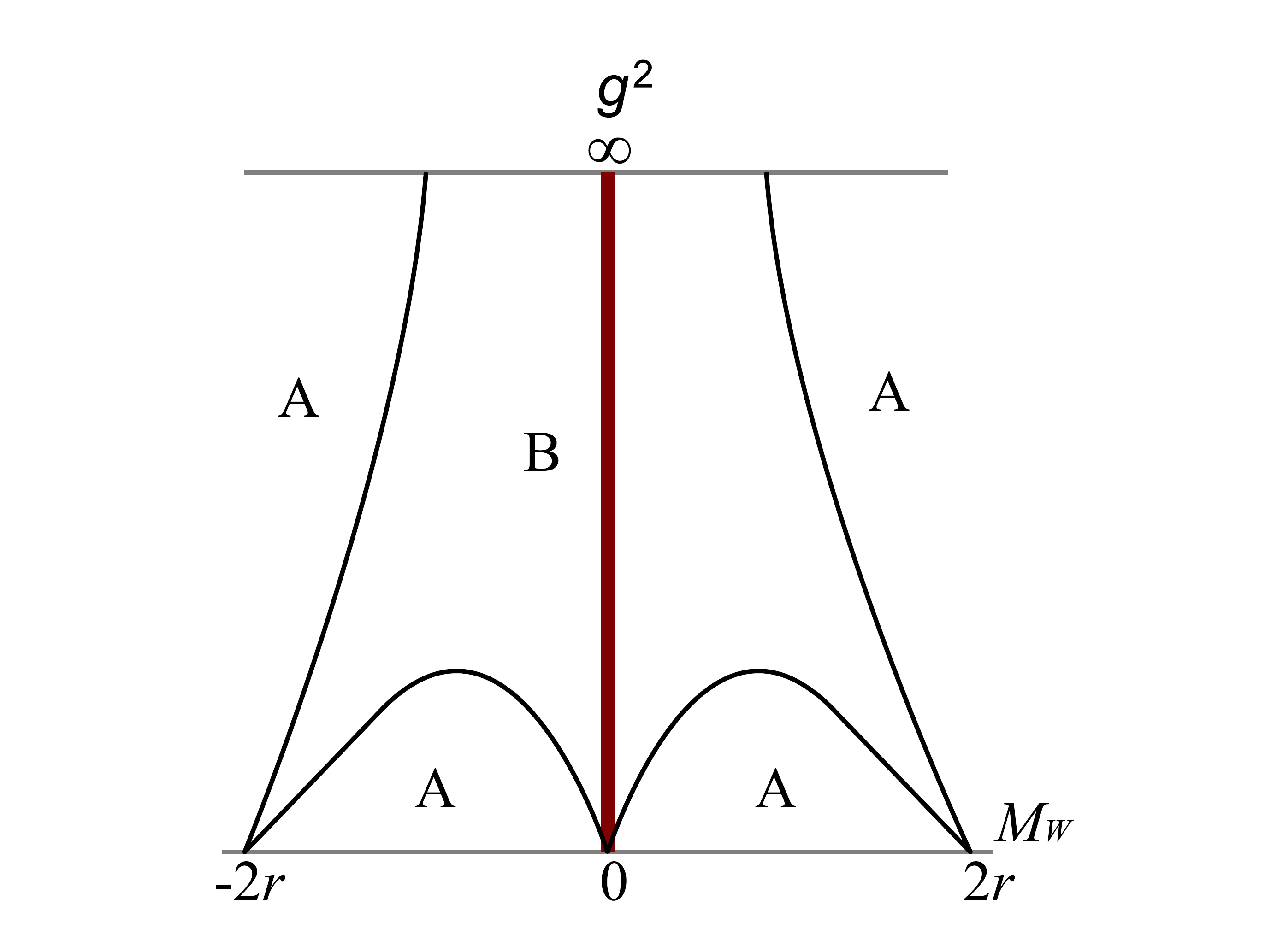} 
\caption{The conjectured 2d Aoki phase diagram in Wilson fermion with four-fermion interaction, which is obtained within the mean-field approximation. Horizontal and vertical axes represent $M_W$ and the four-fermion coupling $g^2$, respectively.  
The phase $A$ is the trivial phase, and the phase $B$ breaks parity by pseudo-scalar condensate $\langle \overline{\psi}\mathrm{i} \gamma_3 \psi \rangle$, which is called Aoki phase. 
The red line corresponds to the central branch, where the extra vector symmetry $U(1)_{\overline{V}}$ emerges, and the central branch is included inside the Aoki phase. 
Our $\mathbb{Z}_2$ 't~Hooft anomaly tells that this central cusp of Aoki phase is the exact result beyond the mean-field approximation. }
\label{WilA}
\end{figure}

The phase structure of Wilson fermion was first studied in Refs.~\cite{Aoki:1983qi, Aoki:1986xr, Aoki:1987us}. 
The $2$d lattice Gross-Neveu model with $N$ Wilson fermions is considered there, and the mean-field gap equation in the large-$N$ limit shows that there is a parity-broken phase due to pseudo-scalar condensate $\langle \overline{\psi}\im \gamma_3 \psi\rangle\not=0$. That parity-broken phase is called Aoki phase. 
The central branch corresponds to the central cusp of the conjectured Aoki phase diagram as shown in Fig.~\ref{WilA}, where the phase $A$ is the trivial one and the phase $B$ is Aoki phase.

In order to establish the connection between our result and Aoki phase, we first translate the results in Refs.~\cite{Aoki:1983qi, Aoki:1986xr, Aoki:1987us} into our setup. We consider the system with four-fermion interaction, 
\be
S=S_{\mathrm{CB}}+M_W\sum_{(x,y)} \overline{\psi}\psi(x,y) +{g^2\over 2}\sum_{(x,y)} \left[\left(\overline{\psi}\psi(x,y)\right)^2+\left(\overline{\psi}\im \gamma_3 \psi(x,y)\right)^2\right]. 
\ee
In order to justify the mean-field gap equation, we have to introduce $N$-flavor lattice fermions and take the large-$N$ limit, but we here just perform the mean-field approximation with $N=1$. 
We then obtain the phase diagram shown in Fig.~\ref{WilA}. 
The central branch is at $M_W=0$, and the mean-field computation shows that there is the pseudo-scalar condensate at any $g^2$. 

We note that the four-fermion coupling explicitly breaks $U(1)_{\overline{V}}$ symmetry at the central branch, because $(\overline{\psi}\psi(x,y))^2\mapsto \rme^{\im 4(-)^{x+y}\beta}(\overline{\psi}\psi(x,y))^2$, etc. However, unlike the mass term, it keeps the non-trivial discrete subgroup, $(\mathbb{Z}_4)_{\overline{V}}$. 
In Sec.~\ref{sec:tHooft_anomaly}, we have shown that there is a $\mathbb{Z}_2$ mixed 't~Hooft anomaly between $(\mathbb{Z}_4)_{\overline{V}}\subset U(1)_{\overline{V}}$, lattice translation, and lattice rotation. 
The pseudo-scalar condensate $\langle \overline{\psi}\im \gamma_3 \psi\rangle$ means the spontaneous symmetry breaking, 
\be
(\mathbb{Z}_4)_{\overline{V}} \xrightarrow{\mathrm{SSB}} (\mathbb{Z}_2)_F, 
\ee
and this means that existence of Aoki phase at any $g^2$ is required by anomaly matching argument. 
If we translate this result into the language of $XYZ$ spin chain, Aoki phase corresponds to the ferromagnetic phase that breaks $(\mathbb{Z}_2\times \mathbb{Z}_2)_{\mathrm{spin}}$ spontaneously. 

Lastly, let us make several remarks before closing this section. 
Since the mean-field computation cannot be justified without the large-$N$ limit, we may have a different phase structure for $N=1$ from Fig.~\ref{WilA}. Whatever it is, we have shown that the 't~Hooft anomaly matching requires that there are at least two degenerate vacua at the central branch $M_W=0$ for any couplings $g^2$. 
Let us again emphasize that the anomaly constraint comes from three symmetries, $(\mathbb{Z}_4)_{\overline{V}}$, lattice translation, and lattice ${\pi\over 2}$ rotation. 
If we break at least one of these symmetries, Aoki phase can terminate at some critical coupling $g_c$, and the system can belong to the trivially gapped phase. In Ref.~\cite{Bermudez:2018eyh}, the phase diagram is studied for anistropic lattices in order to understand the result of density-matrix renormalizatoin group, and it is found that Aoki phase does not extend to the zero coupling when lattice anistropy is introduced.  
This is consistent with our anomaly constraint since lattice ${\pi\over 2}$ rotation is explicitly broken, which clarifies the importance of lattice symmetries. 

In Sec.~\ref{sec:symmetry_spectrum}, we have shown that the central-branch Wilson fermion has no sign problem. 
We have just shown that the $2$d lattice Gross-Neveu model at $M_W=0$ has the pseudo-scalar condensate $\langle \overline{\psi}\im\gamma_3\psi\rangle$ that breaks parity/charge conjugation, so it may cause some questions about the prohibition of parity breaking (or charge-conjugation breaking in $2$d case) by Vafa-Witten theorem~\cite{Vafa:1984xg}. 
Although Vafa-Witten theorem on parity itself has a certain subtlety as discussed in Refs.~\cite{Azcoiti:1999rq, Ji:2001sa}, we note that the Vafa-Witten theorem is circumvented in two ways in the case of this model.  
First, our theorem on semi-positivity assumes that the Dirac operator anti-commutes with $(-)^{x+y}$, but the Hubbard-Stratonovich transformation of the four-fermion coupling violates this assumption. Therefore, the positivity assumption in the Vafa-Witten theorem does not hold, although this still leaves some questions, such as, if we have the sign-problem-free reformulation of the system or not. 
Second, the pseudo-scalar condensate spontaneously breaks $(\mathbb{Z}_4)_{\overline{V}}$ and parity/charge-conjugation separately, but there is a diagonal subgroup that keeps the condensate invariant. 
Using this fact, redefined parity/charge-conjugation with broken internal symmetry is not spontaneously broken. 
We note that the same situation appears in $4$d two-flavor QCD with isospin chemical potential. The conclusion of the Vafa-Witten theorem is evaded as there is no non-zero vacuum expectation value for pseudo-scalar operators that are neutral under other internal symmetries.

\section{Conclusion and discussion}

In this paper, we consider the $2$d lattice $U(1)$ gauge theory with a central-branch Wilson fermion. 
We have shown that this lattice formalism does not have the sign problem, and the numerical Monte Carlo simulation is doable. 
Our proof of semi-positivity of Dirac determinant applies also in four-dimensions, which indicates a new possibility of numerical simulation of multi-flavor QCD on the Wilson central branch. 

Using the low-energy approximation of central-branch Wilson fermion, we have found that the low-energy effective theory has a rich structure of internal symmetries that originates from the lattice symmetry. 
Interestingly, this symmetry has the one-to-one correspondence with that of the Heisenberg $XXZ$ spin chain, and not only the symmetry group but also the 't~Hooft anomaly turn out to be the same. 
The spin rotational symmetry, $SO(2)\rtimes \mathbb{Z}_2$, comes from the site-dependent $U(1)$ symmetry, $U(1)_{\overline{V}}$, and the one-unit lattice translation. 
The lattice translation on $XXZ$ spin chain causes the discrete chiral symmetry in the low-energy effective theory, and the same symmetry is provided by the ${\pi\over 2}$ lattice rotation of our model. 
Let us emphasize that all these symmetries relevant for 't~Hooft anomaly are exact symmetry at the lattice scale. 
Therefore, our lattice formulation realizes the suitable framework for studying the Haldane conjecture as the numerical lattice Monte Carlo simulation. 

 Let us discuss the advantage/disadvantage of the $1$-flavor central-branch Wilson fermion compared with $2$-flavor Wilson fermions for massless $2$-flavor Schwinger model. The advantage of $1$-flavor central-branch fermion is that it has a discrete chiral symmetry, and the nontrivial 't~Hooft anomaly can be realized. This is not possible in the standard Wilson fermion. Disadvantage of central-branch fermion is that its vector-like symmetry is only $O(2)$, so the full $SU(2)/\mathbb{Z}_2$ symmetry is broken. Although we expect that the full vector-like $SU(2)/\mathbb{Z}_2$ symmetry is restored in the continuum limit, more careful study on this point is necessary. 

The central branch corresponds to the central cusp of Aoki phase in $2$d lattice Gross-Neveu model with Wilson fermions. 
Our study of anomaly matching clarifies why there has to be the cusp at the center of Aoki phase. In order to match the 't~Hooft anomaly, there have to be at least two degenerate vacua at the central branch $M_W=0$ for any four-fermion couplings $g^2$. It manifests the existence of Aoki phase for the model beyond the mean-field approximation.
It is a fascinating avenue to extend this study to four dimensions.

We expect that our lattice formulation corresponds to the spacetime lattice discretization of the slave-fermion description of half-integer anti-ferromagnetic spin chain.  
There is also the slave-boson description that leads to the $\mathbb{C}P^1$ sigma model at $\theta=\pi$ in the continuum limit. 
Until recently, there was a difficulty to realize this bosonic description on the $2$d spacetime lattice, but Refs.~\cite{Sulejmanpasic:2019ytl, Gattringer:2018dlw} have shown that there is a lattice discretization having nice consistency with locality and $2\pi$ periodicity of $\theta$ angle, and its dual-variable formulation has no sign problem. 
It can be an interesting future work if we have a nice connection, such as duality, between their and our formulations.


\acknowledgments
The authors thank M.~\"{U}nsal for reading the early draft. 
Y.~T. is supported by JSPS Overseas Research Fellowship. 
The work of T.~M. was in part supported by the Japan Society for the Promotion of Science (JSPS) Grant-in-Aid for Scientific Research (KAKENHI) Grant Numbers 18H01217 and 19K03817.
The work of T.~M. was also supported by the Ministry of Education, Culture, Sports, Science, and Technology(MEXT)-Supported Program for the Strategic Research Foundation at Private Universities ``Topological Science'' (Grant No. S1511006).

\appendix
\section{4d lattice fermions and flavored mass terms}
\label{4d}

In this appendix we summarize the 4d lattice fermions and the possible hyperbubic flavored-mass terms there.
The 4d massless action of naive fermions possesses $U(4)\times U(4)$ flavor-chiral symmetries. In the notation of (\ref{sym_naive_all}), it is generated by
\begin{eqnarray}
\label{4d:sym_t+}
\Gamma^{(+)}_\A&\in &\left\{\mathbf{1}_4\,,\,\, (-1)^{n_1+\ldots+ n_4}\gamma_5\,,\,\,(-1)^{\check{n}_\mu}\gamma_\mu\,,\,\,(-1)^{n_\mu}i \gamma_\mu\gamma_5
\,,\,\,(-1)^{n_{\mu,\nu}}\frac{i \,[\gamma_\mu\,,\gamma_\nu]}{2}\right\}\,,\\
\label{4d:sym_t-}
\Gamma^{(-)}_\A&\in &\left\{(-1)^{n_1+\ldots+ n_4}\mathbf{1}_4\,,\,\, \gamma_5\,,\,\,(-1)^{n_\mu}\gamma_\mu\,,\,\,(-1)^{\check{n}_\mu}i \gamma_\mu\gamma_5
\,,\,\,(-1)^{\check{n}_{\mu,\nu}}\frac{i \,[\gamma_\mu\,,\gamma_\nu]}{2}\right\}\,,
\end{eqnarray}
with $\check{n}_\mu=\sum _{\rho\neq\mu}n_{\rho}$, $n_{\mu,\nu}=n_\mu+n_\nu$ and $\check{n}_{\mu,\nu}=\sum_{\rho\neq\mu,\nu}n_\rho$.
The fermion mass term $\bar{\psi}_{n} \psi_{n}$ breaks this $U(4)\times U(4)$ to the $U(4)$ subgroup $\Gamma^{(+)}_\A$.

The degeneracy of 16 species in naive fermion is lifted and split into 5 branches, 
where 1, 4, 6, 4 and 1 flavors live.
The Wilson term breaks the $U(4)\times U(4)$ invariance to the $U(1)$ invariance under $\mathbf{1}_{4}$ in Eq.~(\ref{4d:sym_t+}). However the special condition $m+4r=0$ leads to the extra invariance under $(-1)^{n_1+\ldots+ n_4}\mathbf{1}_4$ in $\Gamma^{(-)}_{\A}$. 
This condition corresponds to the 4d central-branch Wilson fermion.

We here introduce ``flavored-mass terms'' in four dimensions.
In \cite{Creutz:2010bm}, four nontrivial types of flavored masses are introduced, 
which satisfy $\gamma_{5}$-hermiticity, possess the hypercubic symmetry.
They  can be classified by the number of transporters, 
including the 1-link case as vector (V), 2-link as tensor (T), 
3-link as axial-vector (A) and 4-link as pseudo-scalar (P),
\begin{equation}
 M_{\mathrm{V}} =  \sum_{\mu} C_\mu, \,\,\,\,\,\,\,
 M_{\mathrm{T}} =  \sum_{perm.}\sum_{sym.}C_\mu C_\nu ,\,\,\,\,\,\,\, 
 M_{\mathrm{A}} =  \sum_{perm.}\sum_{sym.}  \prod_{\nu} C_\nu,\,\,\,\,\,\,\,
 M_{\mathrm{P}} =  \sum_{sym.} \prod_{\mu=1}^4 C_\mu,
\label{F-mass}  
\end{equation}
where $\sum_{perm.}$ stands for summation over permutations of the
space-time indices.
$\sum_{perm.}$ and $\sum_{sym.}$ are defined as containing
factors, for example $1/4!$ for $M_{\mathrm{P}}$.
$M_{\mathrm{V}}$ gives the Wilson term as $\sum_{n}\bar{\psi}_{n}(4-M_{\mathrm{V}})\psi_{n}$.
In the momentum space, they are given by $M_{\mathrm{V}}\to\cos p_{\mu}$,
$M_{\mathrm{T}}\to\cos p_{\mu} \cos p_{\nu}$, 
$M_{\mathrm{A}}\to\cos p_{\mu}\cos p_{\nu} \cos p_{\rho}$ 
and $M_{\mathrm{P}}\to\cos p_{1} \cos p_{2} \cos p_{3} \cos p_{4}$.
By use of these terms, we can realize cousins of Wilson fermions.
It is quite notable the two-flavor central-branch fermion is realized by $M_{4{\rm dCB}}=C_{1}+C_{2}+C_{3}+3C_{4}$, which breaks hypercubic symmetry.
Use of the 4d central-branch Wilson fermion in lattice simulations is one of the topics to be further discussed in future.

It is also known that we can introduce flavored-mass terms to staggered fermions, leading to ``staggered-Wilson fermion'' \cite{Adams:2009eb,Adams:2010gx,Hoelbling:2010jw,deForcrand:2012bm,Misumi:2012sp}. The central branch of 4d staggered-Wilson fermion has the enlarged discrete symmetry \cite{Misumi:2012eh}, which prohibits additive mass renormalization. 

\bibliographystyle{utphys}
\bibliography{./QFT,./refs}

\providecommand{\href}[2]{#2}\begingroup\raggedright\begin{thebibliography}{10}

\bibitem{Haldane:1982rj}
F.~D.~M. Haldane, ``{Continuum dynamics of the 1-D Heisenberg antiferromagnetic
  identification with the O(3) nonlinear sigma model},''
\href{http://dx.doi.org/10.1016/0375-9601(83)90631-X}{{\em Phys. Lett.}
  {\bfseries A93} (1983) 464--468}.

\bibitem{Haldane:1983ru}
F.~D.~M. Haldane, ``{Nonlinear field theory of large spin Heisenberg
  antiferromagnets. Semiclassically quantized solitons of the one-dimensional
  easy Axis Neel state},''
\href{http://dx.doi.org/10.1103/PhysRevLett.50.1153}{{\em Phys. Rev. Lett.}
  {\bfseries 50} (1983) 1153--1156}.

\bibitem{Lieb:1961fr}
E.~H. Lieb, T.~Schultz, and D.~Mattis, ``{Two soluble models of an
  antiferromagnetic chain},''
\href{http://dx.doi.org/10.1016/0003-4916(61)90115-4}{{\em Annals Phys.}
  {\bfseries 16} (1961) 407--466}.

\bibitem{Affleck:1986pq}
I.~Affleck and E.~H. Lieb, ``{A Proof of Part of Haldane's Conjecture on Spin
  Chains},''
\href{http://dx.doi.org/10.1007/BF00400304}{{\em Lett. Math. Phys.} {\bfseries
  12} (1986) 57}.

\bibitem{PhysRevLett.84.1535}
M.~Oshikawa, ``Commensurability, excitation gap, and topology in quantum
  many-particle systems on a periodic lattice,''
  \href{http://dx.doi.org/10.1103/PhysRevLett.84.1535}{{\em Phys. Rev. Lett.}
  {\bfseries 84} (Feb, 2000) 1535--1538},
  \href{http://arxiv.org/abs/cond-mat/9911137}{{\ttfamily
  arXiv:cond-mat/9911137 [cond-mat.str-el]}}.

\bibitem{tHooft:1979rat}
G.~'t~Hooft,
  \href{http://dx.doi.org/10.1007/978-1-4684-7571-5_9}{``{Naturalness, chiral
  symmetry, and spontaneous chiral symmetry breaking},''} in {\em {Recent
  Developments in Gauge Theories. Proceedings, Nato Advanced Study Institute,
  Cargese, France, August 26 - September 8, 1979}}, vol.~59, pp.~135--157.
\newblock
1980.
\newblock

\bibitem{Frishman:1980dq}
Y.~Frishman, A.~Schwimmer, T.~Banks, and S.~Yankielowicz, ``{The Axial Anomaly
  and the Bound State Spectrum in Confining Theories},''
\href{http://dx.doi.org/10.1016/0550-3213(81)90268-6}{{\em Nucl. Phys.}
  {\bfseries B177} (1981) 157--171}.

\bibitem{Wen:2013oza}
X.-G. Wen, ``{Classifying gauge anomalies through symmetry-protected trivial
  orders and classifying gravitational anomalies through topological orders},''
  \href{http://dx.doi.org/10.1103/PhysRevD.88.045013}{{\em Phys. Rev.}
  {\bfseries D88} no.~4, (2013) 045013},
\href{http://arxiv.org/abs/1303.1803}{{\ttfamily arXiv:1303.1803 [hep-th]}}.

\bibitem{Kapustin:2014lwa}
A.~Kapustin and R.~Thorngren, ``{Anomalies of discrete symmetries in three
  dimensions and group cohomology},''
  \href{http://dx.doi.org/10.1103/PhysRevLett.112.231602}{{\em Phys. Rev.
  Lett.} {\bfseries 112} no.~23, (2014) 231602},
\href{http://arxiv.org/abs/1403.0617}{{\ttfamily arXiv:1403.0617 [hep-th]}}.

\bibitem{Cho:2014jfa}
G.~Y. Cho, J.~C.~Y. Teo, and S.~Ryu, ``{Conflicting Symmetries in Topologically
  Ordered Surface States of Three-dimensional Bosonic Symmetry Protected
  Topological Phases},''
  \href{http://dx.doi.org/10.1103/PhysRevB.89.235103}{{\em Phys. Rev.}
  {\bfseries B89} no.~23, (2014) 235103},
\href{http://arxiv.org/abs/1403.2018}{{\ttfamily arXiv:1403.2018
  [cond-mat.str-el]}}.

\bibitem{Wang:2014pma}
J.~C. Wang, Z.-C. Gu, and X.-G. Wen, ``{Field theory representation of
  gauge-gravity symmetry-protected topological invariants, group cohomology and
  beyond},'' \href{http://dx.doi.org/10.1103/PhysRevLett.114.031601}{{\em Phys.
  Rev. Lett.} {\bfseries 114} no.~3, (2015) 031601},
\href{http://arxiv.org/abs/1405.7689}{{\ttfamily arXiv:1405.7689
  [cond-mat.str-el]}}.

\bibitem{Witten:2016cio}
E.~Witten, ``{The "Parity" Anomaly On An Unorientable Manifold},''
  \href{http://dx.doi.org/10.1103/PhysRevB.94.195150}{{\em Phys. Rev.}
  {\bfseries B94} no.~19, (2016) 195150},
\href{http://arxiv.org/abs/1605.02391}{{\ttfamily arXiv:1605.02391 [hep-th]}}.

\bibitem{Tachikawa:2016cha}
Y.~Tachikawa and K.~Yonekura, ``{On time-reversal anomaly of 2+1d topological
  phases},'' \href{http://dx.doi.org/10.1093/ptep/ptx010}{{\em PTEP} {\bfseries
  2017} no.~3, (2017) 033B04},
\href{http://arxiv.org/abs/1610.07010}{{\ttfamily arXiv:1610.07010 [hep-th]}}.

\bibitem{Gaiotto:2017yup}
D.~Gaiotto, A.~Kapustin, Z.~Komargodski, and N.~Seiberg, ``{Theta, Time
  Reversal, and Temperature},''
  \href{http://dx.doi.org/10.1007/JHEP05(2017)091}{{\em JHEP} {\bfseries 05}
  (2017) 091},
\href{http://arxiv.org/abs/1703.00501}{{\ttfamily arXiv:1703.00501 [hep-th]}}.

\bibitem{Tanizaki:2017bam}
Y.~Tanizaki and Y.~Kikuchi, ``{Vacuum structure of bifundamental gauge theories
  at finite topological angles},''
  \href{http://dx.doi.org/10.1007/JHEP06(2017)102}{{\em JHEP} {\bfseries 06}
  (2017) 102},
\href{http://arxiv.org/abs/1705.01949}{{\ttfamily arXiv:1705.01949 [hep-th]}}.

\bibitem{Kikuchi:2017pcp}
Y.~Kikuchi and Y.~Tanizaki, ``{Global inconsistency, 't~Hooft anomaly, and
  level crossing in quantum mechanics},''
  \href{http://dx.doi.org/10.1093/ptep/ptx148}{{\em Prog. Theor. Exp. Phys.}
  {\bfseries 2017} (2017) 113B05},
\href{http://arxiv.org/abs/1708.01962}{{\ttfamily arXiv:1708.01962 [hep-th]}}.

\bibitem{Komargodski:2017dmc}
Z.~Komargodski, A.~Sharon, R.~Thorngren, and X.~Zhou, ``{Comments on Abelian
  Higgs Models and Persistent Order},''
\href{http://arxiv.org/abs/1705.04786}{{\ttfamily arXiv:1705.04786 [hep-th]}}.

\bibitem{Komargodski:2017smk}
Z.~Komargodski, T.~Sulejmanpasic, and M.~Unsal, ``{Walls, anomalies, and
  deconfinement in quantum antiferromagnets},''
  \href{http://dx.doi.org/10.1103/PhysRevB.97.054418}{{\em Phys. Rev.}
  {\bfseries B97} no.~5, (2018) 054418},
\href{http://arxiv.org/abs/1706.05731}{{\ttfamily arXiv:1706.05731
  [cond-mat.str-el]}}.

\bibitem{Shimizu:2017asf}
H.~Shimizu and K.~Yonekura, ``{Anomaly constraints on deconfinement and chiral
  phase transition},'' \href{http://dx.doi.org/10.1103/PhysRevD.97.105011}{{\em
  Phys. Rev.} {\bfseries D97} no.~10, (2018) 105011},
\href{http://arxiv.org/abs/1706.06104}{{\ttfamily arXiv:1706.06104 [hep-th]}}.

\bibitem{Wang:2017loc}
J.~Wang, X.-G. Wen, and E.~Witten, ``{Symmetric Gapped Interfaces of SPT and
  SET States: Systematic Constructions},''
  \href{http://dx.doi.org/10.1103/PhysRevX.8.031048}{{\em Phys. Rev.}
  {\bfseries X8} no.~3, (2018) 031048},
\href{http://arxiv.org/abs/1705.06728}{{\ttfamily arXiv:1705.06728
  [cond-mat.str-el]}}.

\bibitem{Gaiotto:2017tne}
D.~Gaiotto, Z.~Komargodski, and N.~Seiberg, ``{Time-reversal breaking in
  QCD$_{4}$, walls, and dualities in 2 + 1 dimensions},''
  \href{http://dx.doi.org/10.1007/JHEP01(2018)110}{{\em JHEP} {\bfseries 01}
  (2018) 110},
\href{http://arxiv.org/abs/1708.06806}{{\ttfamily arXiv:1708.06806 [hep-th]}}.

\bibitem{Tanizaki:2017qhf}
Y.~Tanizaki, T.~Misumi, and N.~Sakai, ``{Circle compactification and 't Hooft
  anomaly},'' \href{http://dx.doi.org/10.1007/JHEP12(2017)056}{{\em JHEP}
  {\bfseries 12} (2017) 056},
\href{http://arxiv.org/abs/1710.08923}{{\ttfamily arXiv:1710.08923 [hep-th]}}.

\bibitem{Tanizaki:2017mtm}
Y.~Tanizaki, Y.~Kikuchi, T.~Misumi, and N.~Sakai, ``{Anomaly matching for phase
  diagram of massless $\mathbb{Z}_N$-QCD},''
  \href{http://dx.doi.org/10.1103/PhysRevD.97.054012}{{\em Phys. Rev.}
  {\bfseries D97} (2018) 054012},
\href{http://arxiv.org/abs/1711.10487}{{\ttfamily arXiv:1711.10487 [hep-th]}}.

\bibitem{Yamazaki:2017dra}
M.~Yamazaki, ``{Relating 't Hooft Anomalies of 4d Pure Yang-Mills and 2d
  $\mathbb{CP}^{N-1}$ Model},''
  \href{http://dx.doi.org/10.1007/JHEP10(2018)172}{{\em JHEP} {\bfseries 10}
  (2018) 172},
\href{http://arxiv.org/abs/1711.04360}{{\ttfamily arXiv:1711.04360 [hep-th]}}.

\bibitem{Guo:2017xex}
M.~Guo, P.~Putrov, and J.~Wang, ``{Time reversal, SU(N) Yang-Mills and
  cobordisms: Interacting topological superconductors/insulators and quantum
  spin liquids in 3+1D},''
  \href{http://dx.doi.org/10.1016/j.aop.2018.04.025}{{\em Annals Phys.}
  {\bfseries 394} (2018) 244--293},
\href{http://arxiv.org/abs/1711.11587}{{\ttfamily arXiv:1711.11587
  [cond-mat.str-el]}}.

\bibitem{Sulejmanpasic:2018upi}
T.~Sulejmanpasic and Y.~Tanizaki, ``{C-P-T anomaly matching in bosonic quantum
  field theory and spin chains},''
  \href{http://dx.doi.org/10.1103/PhysRevB.97.144201}{{\em Phys. Rev.}
  {\bfseries B97} (2018) 144201},
\href{http://arxiv.org/abs/1802.02153}{{\ttfamily arXiv:1802.02153 [hep-th]}}.

\bibitem{Tanizaki:2018xto}
Y.~Tanizaki and T.~Sulejmanpasic, ``{Anomaly and global inconsistency matching:
  $\theta$-angles, $SU(3)/U(1)^2$ nonlinear sigma model, $SU(3)$ chains and its
  generalizations},'' \href{http://dx.doi.org/10.1103/PhysRevB.98.115126}{{\em
  Phys. Rev.} {\bfseries B98} no.~11, (2018) 115126},
\href{http://arxiv.org/abs/1805.11423}{{\ttfamily arXiv:1805.11423
  [cond-mat.str-el]}}.

\bibitem{Yao:2018kel}
Y.~Yao, C.-T. Hsieh, and M.~Oshikawa, ``{Anomaly matching and
  symmetry-protected critical phases in $SU(N)$ spin systems in 1+1
  dimensions},''
\href{http://arxiv.org/abs/1805.06885}{{\ttfamily arXiv:1805.06885
  [cond-mat.str-el]}}.

\bibitem{Kobayashi:2018yuk}
R.~Kobayashi, K.~Shiozaki, Y.~Kikuchi, and S.~Ryu, ``{Lieb-Schultz-Mattis type
  theorem with higher-form symmetry and the quantum dimer models},''
  \href{http://dx.doi.org/10.1103/PhysRevB.99.014402}{{\em Phys. Rev.}
  {\bfseries B99} no.~1, (2019) 014402},
\href{http://arxiv.org/abs/1805.05367}{{\ttfamily arXiv:1805.05367
  [cond-mat.stat-mech]}}.

\bibitem{Tanizaki:2018wtg}
Y.~Tanizaki, ``{Anomaly constraint on massless QCD and the role of Skyrmions in
  chiral symmetry breaking},''
  \href{http://dx.doi.org/10.1007/JHEP08(2018)171}{{\em JHEP} {\bfseries 08}
  (2018) 171},
\href{http://arxiv.org/abs/1807.07666}{{\ttfamily arXiv:1807.07666 [hep-th]}}.

\bibitem{Anber:2018jdf}
M.~M. Anber and E.~Poppitz, ``{Anomaly matching, (axial) Schwinger models, and
  high-T super Yang-Mills domain walls},''
  \href{http://dx.doi.org/10.1007/JHEP09(2018)076}{{\em JHEP} {\bfseries 09}
  (2018) 076},
\href{http://arxiv.org/abs/1807.00093}{{\ttfamily arXiv:1807.00093 [hep-th]}}.

\bibitem{Anber:2018xek}
M.~M. Anber and E.~Poppitz, ``{Domain walls in high-$T SU(N)$ super Yang-Mills
  theory and QCD(adj)},''
\href{http://arxiv.org/abs/1811.10642}{{\ttfamily arXiv:1811.10642 [hep-th]}}.

\bibitem{Armoni:2018bga}
A.~Armoni and S.~Sugimoto, ``{Vacuum structure of charge k two-dimensional QED
  and dynamics of an anti D-string near an O1$^{-}$-plane},''
  \href{http://dx.doi.org/10.1007/JHEP03(2019)175}{{\em JHEP} {\bfseries 03}
  (2019) 175},
\href{http://arxiv.org/abs/1812.10064}{{\ttfamily arXiv:1812.10064 [hep-th]}}.

\bibitem{Hongo:2018rpy}
M.~Hongo, T.~Misumi, and Y.~Tanizaki, ``{Phase structure of the twisted
  $SU(3)/U(1)^2$ flag sigma model on $\mathbb{R}\times S^1$},''
  \href{http://dx.doi.org/10.1007/JHEP02(2019)070}{{\em JHEP} {\bfseries 02}
  (2019) 070},
\href{http://arxiv.org/abs/1812.02259}{{\ttfamily arXiv:1812.02259 [hep-th]}}.

\bibitem{Yonekura:2019vyz}
K.~Yonekura, ``{Anomaly matching in QCD thermal phase transition},''
  \href{http://dx.doi.org/10.1007/JHEP05(2019)062}{{\em JHEP} {\bfseries 05}
  (2019) 062},
\href{http://arxiv.org/abs/1901.08188}{{\ttfamily arXiv:1901.08188 [hep-th]}}.

\bibitem{Nishimura:2019umw}
H.~Nishimura and Y.~Tanizaki, ``{High-temperature domain walls of QCD with
  imaginary chemical potentials},''
  \href{http://dx.doi.org/10.1007/JHEP06(2019)040}{{\em JHEP} {\bfseries 06}
  (2019) 040},
\href{http://arxiv.org/abs/1903.04014}{{\ttfamily arXiv:1903.04014 [hep-th]}}.

\bibitem{Misumi:2019dwq}
T.~Misumi, Y.~Tanizaki, and M.~\"Unsal, ``Fractional $\theta$ angle, 't hooft
  anomaly, and quantum instantons in charge-$q$ multi-flavor schwinger model,''
  \href{http://dx.doi.org/10.1007/JHEP07(2019)018}{{\em JHEP} {\bfseries 07}
  (2019) 018}, \href{http://arxiv.org/abs/1905.05781}{{\ttfamily
  arXiv:1905.05781 [hep-th]}}.

\bibitem{Cherman:2019hbq}
A.~Cherman, T.~Jacobson, Y.~Tanizaki, and M.~\"Unsal, ``{Anomalies, a mod 2
  index, and dynamics of 2d adjoint QCD},''
\href{http://arxiv.org/abs/1908.09858}{{\ttfamily arXiv:1908.09858 [hep-th]}}.

\bibitem{Wilson:1974sk}
K.~G. Wilson, ``{Confinement of Quarks},''
\href{http://dx.doi.org/10.1103/PhysRevD.10.2445}{{\em Phys. Rev.} {\bfseries
  D10} (1974) 2445--2459}.

\bibitem{Creutz:1980zw}
M.~Creutz, ``{Monte Carlo Study of Quantized SU(2) Gauge Theory},''
\href{http://dx.doi.org/10.1103/PhysRevD.21.2308}{{\em Phys. Rev.} {\bfseries
  D21} (1980) 2308--2315}.

\bibitem{Karsten:1980wd}
L.~H. Karsten and J.~Smit, ``{Lattice Fermions: Species Doubling, Chiral
  Invariance, and the Triangle Anomaly},''
  \href{http://dx.doi.org/10.1016/0550-3213(81)90549-6}{{\em Nucl. Phys.}
  {\bfseries B183} (1981) 103}.
[,495(1980)].

\bibitem{Nielsen:1980rz}
H.~B. Nielsen and M.~Ninomiya, ``{Absence of Neutrinos on a Lattice. 1. Proof
  by Homotopy Theory},''
\href{http://dx.doi.org/10.1016/0550-3213(81)90361-8}{{\em Nucl. Phys.}
  {\bfseries B185} (1981) 20}.

\bibitem{Nielsen:1981xu}
H.~B. Nielsen and M.~Ninomiya, ``{Absence of Neutrinos on a Lattice. 2.
  Intuitive Topological Proof},''
\href{http://dx.doi.org/10.1016/0550-3213(81)90524-1}{{\em Nucl. Phys.}
  {\bfseries B193} (1981) 173--194}.

\bibitem{Nielsen:1981hk}
H.~B. Nielsen and M.~Ninomiya, ``{No Go Theorem for Regularizing Chiral
  Fermions},''
\href{http://dx.doi.org/10.1016/0370-2693(81)91026-1}{{\em Phys. Lett.}
  {\bfseries 105B} (1981) 219--223}.

\bibitem{Wilson:1975id}
K.~G. Wilson, \href{http://dx.doi.org/10.1007/978-1-4613-4208-3_6}{``{Quarks
  and Strings on a Lattice},''} in {\em {New Phenomena in Subnuclear Physics:
  Proceedings, International School of Subnuclear Physics, Erice, Sicily, Jul
  11-Aug 1 1975. Part A}}.
\newblock
1975.
\newblock

\bibitem{Kaplan:1992bt}
D.~B. Kaplan, ``{A Method for simulating chiral fermions on the lattice},''
  \href{http://dx.doi.org/10.1016/0370-2693(92)91112-M}{{\em Phys. Lett.}
  {\bfseries B288} (1992) 342--347},
\href{http://arxiv.org/abs/hep-lat/9206013}{{\ttfamily arXiv:hep-lat/9206013
  [hep-lat]}}.

\bibitem{Shamir:1993zy}
Y.~Shamir, ``{Chiral fermions from lattice boundaries},''
  \href{http://dx.doi.org/10.1016/0550-3213(93)90162-I}{{\em Nucl. Phys.}
  {\bfseries B406} (1993) 90--106},
\href{http://arxiv.org/abs/hep-lat/9303005}{{\ttfamily arXiv:hep-lat/9303005
  [hep-lat]}}.

\bibitem{Ginsparg:1981bj}
P.~H. Ginsparg and K.~G. Wilson, ``{A Remnant of Chiral Symmetry on the
  Lattice},''
\href{http://dx.doi.org/10.1103/PhysRevD.25.2649}{{\em Phys. Rev.} {\bfseries
  D25} (1982) 2649}.

\bibitem{Neuberger:1998wv}
H.~Neuberger, ``{More about exactly massless quarks on the lattice},''
  \href{http://dx.doi.org/10.1016/S0370-2693(98)00355-4}{{\em Phys. Lett.}
  {\bfseries B427} (1998) 353--355},
\href{http://arxiv.org/abs/hep-lat/9801031}{{\ttfamily arXiv:hep-lat/9801031
  [hep-lat]}}.

\bibitem{Creutz:2011cd}
M.~Creutz, T.~Kimura, and T.~Misumi, ``{Aoki Phases in the Lattice Gross-Neveu
  Model with Flavored Mass terms},''
  \href{http://dx.doi.org/10.1103/PhysRevD.83.094506}{{\em Phys. Rev.}
  {\bfseries D83} (2011) 094506},
\href{http://arxiv.org/abs/1101.4239}{{\ttfamily arXiv:1101.4239 [hep-lat]}}.

\bibitem{Kimura:2011ik}
T.~Kimura, S.~Komatsu, T.~Misumi, T.~Noumi, S.~Torii, and S.~Aoki,
  ``{Revisiting symmetries of lattice fermions via spin-flavor
  representation},'' \href{http://dx.doi.org/10.1007/JHEP01(2012)048}{{\em
  JHEP} {\bfseries 01} (2012) 048},
\href{http://arxiv.org/abs/1111.0402}{{\ttfamily arXiv:1111.0402 [hep-lat]}}.

\bibitem{Misumi:2012eh}
T.~Misumi, ``{New fermion discretizations and their applications},''
  \href{http://dx.doi.org/10.22323/1.164.0005}{{\em PoS} {\bfseries
  LATTICE2012} (2012) 005},
\href{http://arxiv.org/abs/1211.6999}{{\ttfamily arXiv:1211.6999 [hep-lat]}}.

\bibitem{Chowdhury:2013ux}
A.~Chowdhury, A.~Harindranath, J.~Maiti, and S.~Mondal, ``Many avatars of the
  wilson fermion: A perturbative analysis,''
  \href{http://dx.doi.org/10.1007/JHEP02(2013)037}{{\em JHEP} {\bfseries 02}
  (2013) 037}, \href{http://arxiv.org/abs/1301.0675}{{\ttfamily arXiv:1301.0675
  [hep-lat]}}.

\bibitem{Creutz:2010bm}
M.~Creutz, T.~Kimura, and T.~Misumi, ``{Index Theorem and Overlap Formalism
  with Naive and Minimally Doubled Fermions},''
  \href{http://dx.doi.org/10.1007/JHEP12(2010)041}{{\em JHEP} {\bfseries 12}
  (2010) 041},
\href{http://arxiv.org/abs/1011.0761}{{\ttfamily arXiv:1011.0761 [hep-lat]}}.

\bibitem{M1}
T.~Misumi. {PhD thesis, Kyoto University}, 2012.
\newblock http://hdl.handle.net/2433/157773.

\bibitem{Adler:1969gk}
S.~L. Adler, ``{Axial vector vertex in spinor electrodynamics},''
\href{http://dx.doi.org/10.1103/PhysRev.177.2426}{{\em Phys. Rev.} {\bfseries
  177} (1969) 2426--2438}.

\bibitem{Bell:1969ts}
J.~S. Bell and R.~Jackiw, ``{A PCAC puzzle: pi0 --> gamma gamma in the sigma
  model},''
\href{http://dx.doi.org/10.1007/BF02823296}{{\em Nuovo Cim.} {\bfseries A60}
  (1969) 47--61}.

\bibitem{Pantev:2005zs}
T.~Pantev and E.~Sharpe, ``{GLSM's for Gerbes (and other toric stacks)},''
  \href{http://dx.doi.org/10.4310/ATMP.2006.v10.n1.a4}{{\em Adv. Theor. Math.
  Phys.} {\bfseries 10} no.~1, (2006) 77--121},
\href{http://arxiv.org/abs/hep-th/0502053}{{\ttfamily arXiv:hep-th/0502053
  [hep-th]}}.

\bibitem{Pantev:2005wj}
T.~Pantev and E.~Sharpe, ``{String compactifications on Calabi-Yau stacks},''
  \href{http://dx.doi.org/10.1016/j.nuclphysb.2005.10.035}{{\em Nucl. Phys.}
  {\bfseries B733} (2006) 233--296},
\href{http://arxiv.org/abs/hep-th/0502044}{{\ttfamily arXiv:hep-th/0502044
  [hep-th]}}.

\bibitem{Aoki:1983qi}
S.~Aoki, ``{New Phase Structure for Lattice QCD with Wilson Fermions},''
\href{http://dx.doi.org/10.1103/PhysRevD.30.2653}{{\em Phys. Rev.} {\bfseries
  D30} (1984) 2653}.

\bibitem{Aoki:1986xr}
S.~Aoki, ``{A Solution to the U(1) Problem on a Lattice},''
\href{http://dx.doi.org/10.1103/PhysRevLett.57.3136}{{\em Phys. Rev. Lett.}
  {\bfseries 57} (1986) 3136}.

\bibitem{Aoki:1987us}
S.~Aoki, ``{U(1) Problem and Lattice QCD},''
\href{http://dx.doi.org/10.1016/0550-3213(89)90113-2}{{\em Nucl. Phys.}
  {\bfseries B314} (1989) 79--111}.

\bibitem{takahashi1999thermodynamics}
M.~Takahashi, {\em Thermodynamics of one-dimensional solvable models}.
\newblock Cambridge University Press, 1999.

\bibitem{Bermudez:2018eyh}
A.~Bermudez, E.~Tirrito, M.~Rizzi, M.~Lewenstein, and S.~Hands,
  ``{Gross--Neveu--Wilson model and correlated symmetry-protected topological
  phases},'' \href{http://dx.doi.org/10.1016/j.aop.2018.10.007}{{\em Annals
  Phys.} {\bfseries 399} (2018) 149--180},
\href{http://arxiv.org/abs/1807.03202}{{\ttfamily arXiv:1807.03202
  [cond-mat.quant-gas]}}.

\bibitem{Vafa:1984xg}
C.~Vafa and E.~Witten, ``{Parity Conservation in QCD},''
\href{http://dx.doi.org/10.1103/PhysRevLett.53.535}{{\em Phys. Rev. Lett.}
  {\bfseries 53} (1984) 535}.

\bibitem{Azcoiti:1999rq}
V.~Azcoiti and A.~Galante, ``{Parity and CT realization in QCD},''
  \href{http://dx.doi.org/10.1103/PhysRevLett.83.1518}{{\em Phys. Rev. Lett.}
  {\bfseries 83} (1999) 1518--1520},
\href{http://arxiv.org/abs/hep-th/9901068}{{\ttfamily arXiv:hep-th/9901068
  [hep-th]}}.

\bibitem{Ji:2001sa}
X.-d. Ji, ``{Validity of the Vafa-Witten proof on absence of spontaneous parity
  breaking in QCD},''
  \href{http://dx.doi.org/10.1016/S0370-2693(02)03273-2}{{\em Phys. Lett.}
  {\bfseries B554} (2003) 33--37},
\href{http://arxiv.org/abs/hep-ph/0108162}{{\ttfamily arXiv:hep-ph/0108162
  [hep-ph]}}.

\bibitem{Sulejmanpasic:2019ytl}
T.~Sulejmanpasic and C.~Gattringer, ``{Abelian gauge theories on the lattice:
  $\theta$-terms and compact gauge theory with(out) monopoles},''
  \href{http://dx.doi.org/10.1016/j.nuclphysb.2019.114616}{{\em Nucl. Phys.}
  {\bfseries B943} (2019) 114616},
\href{http://arxiv.org/abs/1901.02637}{{\ttfamily arXiv:1901.02637 [hep-lat]}}.

\bibitem{Gattringer:2018dlw}
C.~Gattringer, D.~G\"{o}schl, and T.~Sulejmanpasic, ``{Dual simulation of the
  2d U(1) gauge Higgs model at topological angle $\theta = \pi\,$: Critical
  endpoint behavior},''
  \href{http://dx.doi.org/10.1016/j.nuclphysb.2018.08.017}{{\em Nucl. Phys.}
  {\bfseries B935} (2018) 344--364},
\href{http://arxiv.org/abs/1807.07793}{{\ttfamily arXiv:1807.07793 [hep-lat]}}.

\bibitem{Adams:2009eb}
D.~H. Adams, ``Theoretical foundation for the index theorem on the lattice with
  staggered fermions,''
  \href{http://dx.doi.org/10.1103/PhysRevLett.104.141602}{{\em Phys.Rev.Lett.}
  {\bfseries 104} (2010) 141602},
  \href{http://arxiv.org/abs/0912.2850}{{\ttfamily arXiv:0912.2850 [hep-lat]}}.

\bibitem{Adams:2010gx}
D.~H. Adams, ``Pairs of chiral quarks on the lattice from staggered fermions,''
  \href{http://dx.doi.org/10.1016/j.physletb.2011.04.034}{{\em Phys.Lett.B}
  {\bfseries 699} (2011) 394--397},
  \href{http://arxiv.org/abs/1008.2833}{{\ttfamily arXiv:1008.2833 [hep-lat]}}.

\bibitem{Hoelbling:2010jw}
C.~Hoelbling, ``Single flavor staggered fermions,''
  \href{http://dx.doi.org/10.1016/j.physletb.2010.12.062}{{\em Phys.Lett.B}
  {\bfseries 696} (2011) 422--425},
  \href{http://arxiv.org/abs/1009.5362}{{\ttfamily arXiv:1009.5362 [hep-lat]}}.

\bibitem{deForcrand:2012bm}
P.~de~Forcrand, A.~Kurkela, and M.~Panero, ``Numerical properties of staggered
  quarks with a taste-dependent mass term,''
  \href{http://dx.doi.org/10.1007/JHEP04(2012)142}{{\em JHEP} {\bfseries 04}
  (2012) 142}, \href{http://arxiv.org/abs/1202.1867}{{\ttfamily arXiv:1202.1867
  [hep-lat]}}.

\bibitem{Misumi:2012sp}
T.~Misumi, T.~Z. Nakano, T.~Kimura, and A.~Ohnishi, ``Strong-coupling analysis
  of parity phase structure in staggered-wilson fermions,''
  \href{http://dx.doi.org/10.1103/PhysRevD.86.034501}{{\em Phys.Rev.D}
  {\bfseries 86} (2012) 034501},
  \href{http://arxiv.org/abs/1205.6545}{{\ttfamily arXiv:1205.6545 [hep-lat]}}.

\end{thebibliography}\endgroup
\end{document}